\documentclass[a4paper,fleqn,usenatbib]{mnras}

\usepackage{graphics}
\usepackage{graphicx}
\usepackage{psfig}
\input{epsf}
\usepackage{amsmath}
\usepackage{amssymb}
\usepackage{url}
\usepackage[T1]{fontenc}
\usepackage{ae,aecompl}
\usepackage{widetext}

\newcommand{\be}{\begin{equation}}
\newcommand{\ee}{\end{equation}}

\newcommand{\order}[1]{\mathcal{O}\!\left(#1\right)}
\newcommand{\uS}{\mathrm{s}}
\newcommand{\uc}{\mathrm{c}}
\newcommand{\ud}{\mathrm{d}}
\newcommand{\bfmath}[1]{\ensuremath{\boldsymbol{#1}}}

\title[Multi-scale Pipeline for Cosmic String] {Multi-Scale Pipeline for
  the Search of String-Induced CMB Anisotropies}
\author[Vafaei et al.]{
A. Vafaei Sadr$^{1,2,3}$, S. M. S. Movahed$^{1,4}$\thanks{E-mail: m.s.movahed@ipm.ir}, M. Farhang$^{1}$, C. Ringeval$^{5}$, F. R. Bouchet$^{6}$
\\
$^{1}$Department of Physics, Shahid Beheshti University,
Velenjak, Tehran 19839, Iran\\
$^{2}$D\'epartement de Physique Th\'eorique and Center for Astroparticle Physics, Universit\'e de Genève, 24 Quai Ernest Ansermet,\\ 1211 Gen\'eve 4, Switzerland\\
$^{3}$African Institute for Mathematical Sciences, 6 Melrose Road, Muizenberg, 7945, South Africa\\
$^{4}$ School of Physics, Institute for Research in Fundamental Sciences (IPM), P. O. Box 19395-5531, Tehran, Iran\\
$^{5}$Centre for Cosmology, Particle Physics and Phenomenology, Universit\'e Catholique de Louvain, Louvain-la-Neuve B-1348, Belgium\\
$^{6}$Institut d' Astrophysique de Paris (UMR7095: CNRS \& UPMC-Sorbonne Universities), F-75014, Paris, France}

\begin{document}
\maketitle

\begin{abstract}
We propose a multi-scale edge-detection algorithm to search for the
Gott-Kaiser-Stebbins imprints of a cosmic string (CS) network on the
Cosmic Microwave Background (CMB) anisotropies. Curvelet
decomposition and extended Canny algorithm are used to enhance the
string detectability. Various statistical tools are then applied to
quantify the deviation of CMB maps having a cosmic string
contribution with respect to pure Gaussian anisotropies
of inflationary origin. These statistical measures include the
one-point probability density function, the weighted two-point
correlation function (TPCF) of the anisotropies, the unweighted TPCF
of the peaks and of the up-crossing map, as well as their
cross-correlation.
We use this algorithm on a hundred of
  simulated Nambu-Goto CMB flat sky maps, covering approximately $10\%$ of the sky, and for different string
  tensions $G\mu$.  On noiseless sky maps with an angular
  resolution of $0.9'$, we show that our pipeline detects CSs with $G\mu$ as low
  as $G\mu\gtrsim 4.3\times 10^{-10}$. At the same resolution, but with a
  noise level typical to a CMB-S4 phase II experiment, the detection threshold
  would be to $G\mu\gtrsim 1.2 \times 10^{-7}$.

\end{abstract}

\begin{keywords}
cosmic background radiation - cosmology; theory - early Universe -
large-scale structure of Universe.
\end{keywords}

\section{Introduction}
The inflationary $\Lambda$CDM model with nearly Gaussian and scale-invariant primordial density perturbations has been confirmed with high precision as a robust cosmological model thanks in particular to the observations of the cosmic microwave background radiation (CMB) \citep{Hinshaw:2012aka,Ade:2015xua}. The initial conditions for the large-scale structure of the Universe, determined by primordial cosmological perturbations are seeded by quantum fluctuations of a scalar field during the so-called inflationary epoch \citep{Guth:1980zm,Liddle:1993fq,Steinhardt:1995uf,Liddle:1999mq}.
Despite the outstanding agreement between the standard model and the cosmic data, there is some limited  room for alternative scenarios as well. One such scenario is to consider topological defects as minor contributors to the primordial perturbations.
Many quantum filed theories typically predict these defects as a result of phase transition caused by spontaneous breaking of their symmetries due to the expansion and cooling of the Universe \citep{Kibble:1976sj,Kibble:1980mv22,Hindmarsh:1994re,Vilenkin:2000jqa,Copeland:2009ga,Polchinski:2004hb}.

The line-like version of topological defects are called cosmic strings (CSs) and are commonly present in theories of
hybrid inflation, brane-world models and superstring theory \citep{Kibble:1976sj,Zeldovich:1980gh,Vilenkin:1981iu,
Vachaspati:1984dz,Vilenkin:1984ib,Shellard:1987bv,Hindmarsh:1994re,Vilenkin:2000jqa,Sakellariadou:2006qs,Bevis:2007gh,Depies:2009im,Bevis:2010gj,Copeland:1994vg,Sakellariadou:1997zt,Sarangi:2002yt,Copeland:2003bj,Pogosian:2003mz,Majumdar:2002hy,Dvali:2003zj,Kibble:2004hq,HenryTye:2006uv}.
They represent lines of trapped energy density
parameterized by $G\mu$. $G$ is Newton's constant and $\mu$
represents the mass per unit length of the string, also equal to its
tension. The string tension is closely related to the energy of the
symmetry breaking scale, $\varpi$, as:
\begin{equation}
\frac{G\mu}{c^2}=\order{\frac{\varpi^2}{M_{\rm Planck}^2}}.
\end{equation}
here $M_{\rm Planck}\equiv\sqrt{\hbar c/G}$ represents the Planck's mass and $c$ is the speed of light. In this paper we choose to work in natural units with $\hbar=c=1$.
Symmetry breaking at energies around the GUT scale would thus correspond to production of CSs with  $G\mu \sim
10^{-6}$ \citep{Kibble:1976sj,Zeldovich:1980gh,Vilenkin:1981iu,Vilenkin:2000jqa,Firouzjahi:2005dh}.
 Therefore, CS studies provide a unique path to the physics of extremely high energies far beyond the access of our Earth-bound laboratories.
The evolution of a network of CSs, containing loops, long strings and
their junctions,  depends not only on the string tension, but also on
the equation of motion of the strings, the initial conditions and the
string inter-commutation probability, that represents the probability
of their collisions,  \citep{Vachaspati:1984dz,Ringeval:2005kr, BlancoPillado:2011dq}.

The search for CSs takes different theoretical, statistical and
observational routes, thanks to their diverse imprints on cosmological
data sets. These searches have led to constraints on $G\mu$, which is
the main free parameter characterizing CSs. For example, recent
  results from the gravitational wave emission of Nambu-Goto cosmic
  string loops constrain the CS tension to be $10^{-14} \le G\mu \le
  1.5 \times 10^{-10}$ depending on the string microstructure~\citep{Ringeval:2017eww, Blanco-Pillado:2017oxo,Blanco-Pillado:2017rnf}. Pulsar timing and photometry, based on gravitational microlensing, constrain CS's tension to
$10^{-15}<G\mu<10^{-8}$~\citep{Jenet:2006sv,Pshirkov:2009vb,Tuntsov:2010fu, Damour:2004kw,Battye:2010xz,Oknyanskij:2005pd, Kuroyanagi:2012jf}.
 The upper bound of $G\mu< 3\times 10^{-7}$  has also been reported by the COSMOS survey \citep{Christiansen:2010zi}.
 The $21$-cm signature of CS wakes has also been theoretically
 explored in \cite{Brandenberger:2010hn, Hernandez:2011ym, Hernandez:2012qs,Pagano:2012cx} and forecasts have been made on how strongly these near-future surveys would measure $G\mu$. On the other hand, \cite{Shlaer:2012rj}  have studied  signature of CSs on high-redshift large-scale structure surveys and on the ionization history of the Universe.

The CS network, if it exists, should have also left unique imprints
on CMB anisotropies. Their integrated Sachs-Wolfe (ISW) contribution,
also known as the Gott-Kaiser-Stebbins effect~\citep{Kaiser:1984iv,
  Gott:1985, Stebbins:1988, Bouchet:1988hh, Allen:1997ag, Pen:1997ae},
is primarily caused by the transverse motion of the CSs with respect
to the observer. The resulting energy shift of CMB photons produces
line-like discontinuities on CMB anisotropies at the string location
such that, in the light-cone gauge, one has~\citep{Kaiser:1984iv,
  Gott:1985, Hindmarsh:1993pu, Stebbins:1995}
\begin{equation}
    \frac{\delta T}{T} \sim 8\pi
    G\mu v_{\uS},
\end{equation}
where $v_\uS$ is the transverse velocity of the string.

Simulating the impact of the CS network on CMB anisotropies requires
various simplifying assumptions. The models used in the literature
generally fall in one of the followings:
 \\ $(i)$ Nambu-Goto simulations
\citep{Bennett:1987vf,Bennett:1989ak, Bouchet:1988hh, Landriau:2002fx,Landriau:2003xf, Fraisse:2007nu, Ringeval:2012tk}, $(ii)$ using
stochastic ensemble of unconnected segments \citep{Allen:1997ag,Albrecht:1997nt, Contaldi:1998mx, Pogosian:1999np, Pogosian:2006hg},
$(iii)$ Abelian-Higgs model on a lattice, with the evolution of the
network determined by the corresponding fields
\citep{Vincent:1997cx,Moore:2001px,Kasuya:1999hy,Bevis:2006mj}, and
$(iv)$ the so-called statistical approach, explained below
\citep{Perivolaropoulos:1992gy, Perivolaropoulos:1992if,Moessner:1993za, Jeong:2004ut, Amsel:2007ki, Stewart:2008zq,Danos:2008fq, Movahed:2010zq}.

The $(i)-(iii)$ approaches solve the photon propagation at linear
order within a CS simulation network to get CMB fluctuations.
Previous results from these models showed that at intermediate and
small scales, topological defects and inflationary models lead to
completely different results, while at large enough scales both
scenarios result in similar features in the CMB power spectrum.
The fourth method uses the number of random kicks on photon
trajectories by CSs network between the time of recombination and the
present era. This approach requires dealing with analytical and
numerical tools and is explained in detail in \cite{Stewart:2008zq,Danos:2008fq,Movahed:2010zq}.

To measure the contribution of CSs to the CMB power spectrum, the
standard parameter estimation techniques are extended to include a new
parameter, usually denoted by $f_{10}$, quantifying the fraction of
the power at $\ell=10$ due to strings. The incoherency of the
perturbations produced by the strings as active sources leads to a
significantly broad peak compared to the relatively sharp peak from the standard
acoustic oscillations. The measurements of the CMB power spectrum
leave limited space for contribution from CS-induced perturbations
\citep{Pen:1997ae,Bevis:2006mj,Bevis:2010gj,Lazanu:2014eya}.
The {\it Planck} data \citep{Ade:2013xla} constrains this contribution to
be $f_{10}<0.024$ (corresponding to $G\mu<3.0\times10^{-7}$) for
Abelian-Higgs strings and $f_{10}<0.010$ (corresponding to
$G\mu<1.3\times10^{-7}$) for unconnected segments. Adding CMB
polarization improves the upper bound to $G\mu<1.1 \times
10^{-7}$~\citep{Charnock:2016nzm}. Latest results for Nambu-Goto
strings give an upper bound of $G \mu < 1.5 \times 10^{-7}$ from the
{\it Planck} data with polarization~\citep{Lazanu:2014eya}. One should note though that {\it Planck} 15 polarization data is preliminary
at large multipoles due to residual systematics at the $\mathcal {O}(1\mu K^2)$ level.

An alternative approach to constrain $G\mu$
is based on the non-Gaussianity of CS-induced
fluctuations~\citep{Ringeval:2010ca, Ducout:2012it}. For example, bispectrum
measurement of the observed CMB anisotropies, Wavelet-based data
analysis methods and measurements of the Minkowski functionals of the
CMB data have set the upper bounds of $G\mu<8.8\times10^{-7}$,
$G\mu<7\times10^{-7}$ and $G\mu<7.8\times10^{-7}$, respectively
\citep{Hindmarsh:2009qk, Hindmarsh:2009es, Ade:2013xla, Regan:2015cfa}.

On the other hand, the discontinuities on the CMB anisotropy maps
produced by the CS network most clearly manifest themselves in the
real-space approaches. These methods are expected to be less time
consuming compared to Fourier-based approaches. Among the real-space
methods with strings modeled as random kicks  \citep{Movahed:2010zq}
used the crossing statistics of simulated ideal CMB fluctuations and claimed detectability of CSs with
 $G\mu\gtrsim 4.0\times 10^{-9}$.
Using the unweighted Two-Point Correlation Function (TPCF) of CMB
peaks instead increases the detectability threshold to $G\mu\gtrsim 1.2\times 10^{-8}$ for noiseless, 1'-resolution maps \citep{Movahed:2012zt}.

  Another potentially powerful method in real-space analysis is
  to exploit our knowledge of the anisotropy patterns from CSs, i.e.,
  the line-like edges. \citep{Stewart:2008zq} applied edge-detection
  algorithms on ideal random-kick  maps
  to get a detection threshold of $G\mu \gtrsim
  5.5\times 10^{-8}$ for the South Pole Telescope (SPT). Using wavelet
  and curvelet methods, \citep{Hergt:2016xup} found a sensitivity of
  $G\mu\gtrsim 1.4 \times 10^{-7}$ for the SPT third generation.
  Recently, neural network-based approaches have been applied by
  \citep{Ciuca:2017wrk} on
  noiseless arcminute-resolution random-kick maps to reach a detection threshold
  of $G\mu \gtrsim 5 \times 10^{-9}$.

  The different values reported above are, in part, the
  results of the crude assumptions made to model the strings. For this
  reason, in the following, we will be using small angle CMB maps
  directly computed from Nambu-Goto
  simulations~\citep{Ringeval:2005kr, Fraisse:2007nu,
    Ringeval:2012tk}. In particular, these maps are the flat version
  of the ones recently used by \citep{McEwen:2016aob} which reported a
  Bayesian detection threshold of $G\mu\sim 5\times 10^{-7}$ for a
  {\it Planck}-like CMB experiment. This allows a fair comparison with
  our results for the {\it Planck}-like CMB maps (see
  section~\ref{sec:results}). For forthcoming arcminute-resolution
  experiments \citep{Hammond:2008fg} used wavelet-domain Bayesian
  denoising on two of the maps we have used to obtain the detection lower bound
  of $G\mu\ge 1.0\times 10^{-7}$.

In this work, we develop a new pipeline to search for the CS
signals. Taking the CMB map as the input, the pipeline follows several
image processing steps to enhance the detectability of the CS
trace. More specifically, the CMB map is decomposed into various
curvelet components, with different scales, so that only components
with the highest contribution from CSs are kept for further analysis.
These components are then passed through certain filters to produce
gradient maps thereby boosting the CS-induced
discontinuities. At the last step of the pipeline, various statistical
measures are applied on the gradient maps to quantify possible
deviations from inflation-induced anisotropies.

%

Our proposed pipeline has certain degrees of freedom which set its
adjustable parameters such as the curvelet component to be used, the
filter type in the edge-detection step, and the kind of the
statistical measure to be applied on the gradient maps.  For each
experimental setup, the pipeline automatically searches for the
optimum sequence of parameters yielding the tightest constraint on the
CS contribution.

The outline of this paper is as follows. In Section 2, we introduce
the different components of our CMB simulations.  Section 3 describes
in details our proposed pipeline for CS detection, and in Section 4,
we present the performance of the pipeline by applying it on simulated
CMB data with various noise levels. We conclude in Section 5.

\section{Simulation of CMB maps}
 In this section, we describe the details of our simulations for making CMB sky maps, used in Section \ref{sec:results} to investigate the detectability of the string contribution to the CMB anisotropies. The simulations consist of three components: (1) the Gaussian inflation-induced contribution denoted by $G$, as well as the secondary lensing signal (Section~\ref{sec:Gsignal}), (2) the CS contribution, $G\mu \times S$, where $S$  represents the  normalized simulated template for the string signal and $G\mu$ sets its amplitude (Section~\ref{sec:cs}), and (3) the experimental noise indicated by $N$ (section~\ref{sec:Gnoise}).
 The full simulated map  $T(x,y)$, with $x$ and $y$ representing pixel coordinates, would then be
\be\label{fullmap}
T(x,y)=B\left[G(x,y)+G\mu \times S(x,y)\right]+N(x,y).
\ee
 where $B$ characterizes the beam function  (Section~\ref{sec:beam}).  In the figures throughout this paper, we use G for Gaussian simulated map, S for the CS-induced anisotropy map and N  for the noise map.
 For example, GSBN refers to simulations with all components included, with B representing the beam effect.  
  We work in the flat sky limit
   \citep{Heavens:1999cq} with $100$ square maps of side
   $\Theta=7.2^\circ$, with $1024\times 1024$ pixels. This corresponds
   to a resolution of $R=0.42'$ before convolution with an
   experimental beam. We also assume
 statistical isotropy in all our simulations.

\begin{figure}
\begin{center}
\includegraphics[scale=0.4]{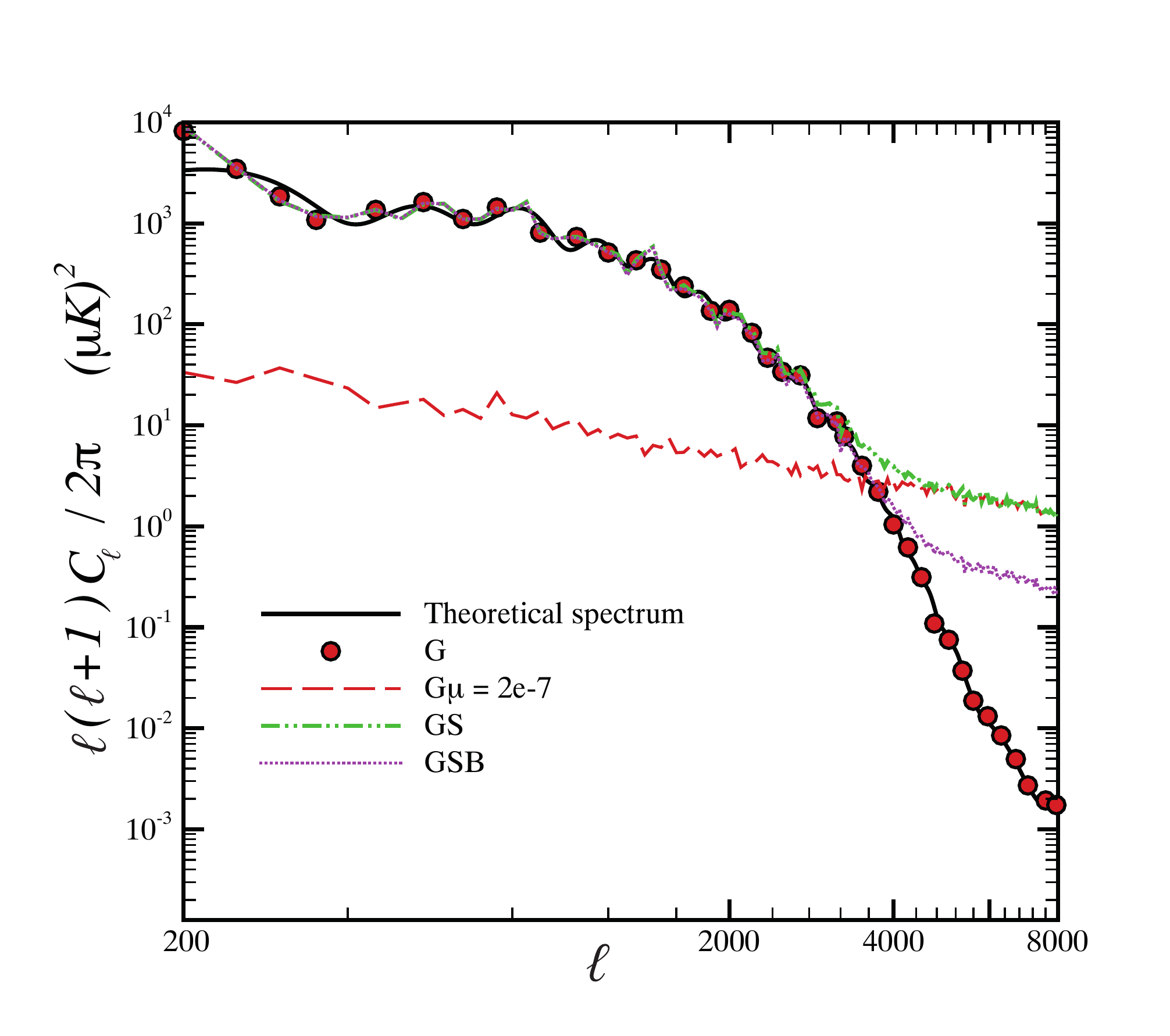}
\end{center}
\caption{The CMB power spectrum: the black solid line shows the
  fiducial power spectrum for the
   $\Lambda$CDM model  compatible with {\it Planck} 15 \citep{Aghanim:2015xee}  (computed by CAMB).
  The red filled circles represent the measured power spectrum of simulated
  Gaussian maps seeded only by inflationary fluctuations, with pixel
  resolution of $R=0.42'$ and map size of $\Theta=7.2^{\circ}$.
The red long-dashed line shows the contribution to the power spectrum from CS network characterized by $G\mu=2.0\times10^{-7}$. The dashed-dot-dot curve corresponds to the measured power spectrum of  CMB maps including both inflationary and CS-induced anisotropies.
The dotted curve corresponds to the measured power spectrum of maps smeared by the beam (Section~\ref{sec:beam}).  One sees that, in the absence of noise and other small scale contaminations, the CS component is most easily detected at $\ell\gtrsim 4000$.}
\label{fig:power}
\end{figure}
%

\subsection{Gaussian CMB simulation}\label{sec:Gsignal}

The Gaussian component of CMB temperature anisotropies is assumed to be seeded by adiabatic scale-invariant slow-roll-inflationary fluctuations.  The only secondary contribution considered here is due to lensing. %
To this end, we use the CAMB
software\footnote{\texttt{http://camb.info}} \citep{Lewis:1999bs} to
calculate the temperature power spectrum for the parameter set of
the $\Lambda$CDM model consistent with {\it Planck} 15, Supernova
type Ia (SNIa) and the Sloan Digital Sky Survey (SDSS) data sets
\citep{Aghanim:2015xee}.  The computed $C_\ell $ will be used to
generate 2D Gaussian random fields following \citep{Bond:1987ub}.
 The maps $G(x,y)$ are generated by Fourier-transforming Gaussian random realizations $G(\mathbf k)$ of CMB temperature  power spectrum in the flat sky limit $P_{TT}(k)$,
\be
G(\mathbf k) = \sqrt{\frac{P_{ TT}(k)}{2}}(\mathcal{R}_1+i\mathcal{R}_2),
\ee
where  $\mathcal{R}_1$ and $\mathcal{R}_2$ are two mean-less unit variance normal random fields, and $k=|\mathbf{k}|$. For the flat power spectrum $P_{TT}(k)$ one has  $\langle {\delta_{T}}({\mathbf k}){\delta_{T}}^*({\mathbf k}')\rangle=(2\pi)^2P_{TT}(k)\delta_{\rm d}({\mathbf k}-{\mathbf k}')$, where $\delta_{\rm d}$ denotes the Dirac delta function.
$P_{TT}(k)$ is related to the full sky power spectrum through  $\ell(\ell+1)C_{\ell}^{TT}\sim k^2P_{TT}( k)$ \citep {White:1997wq,Hindmarsh:2009qk,Fraisse:2007nu}.

\begin{figure}
\begin{center}
\includegraphics[scale=0.4]{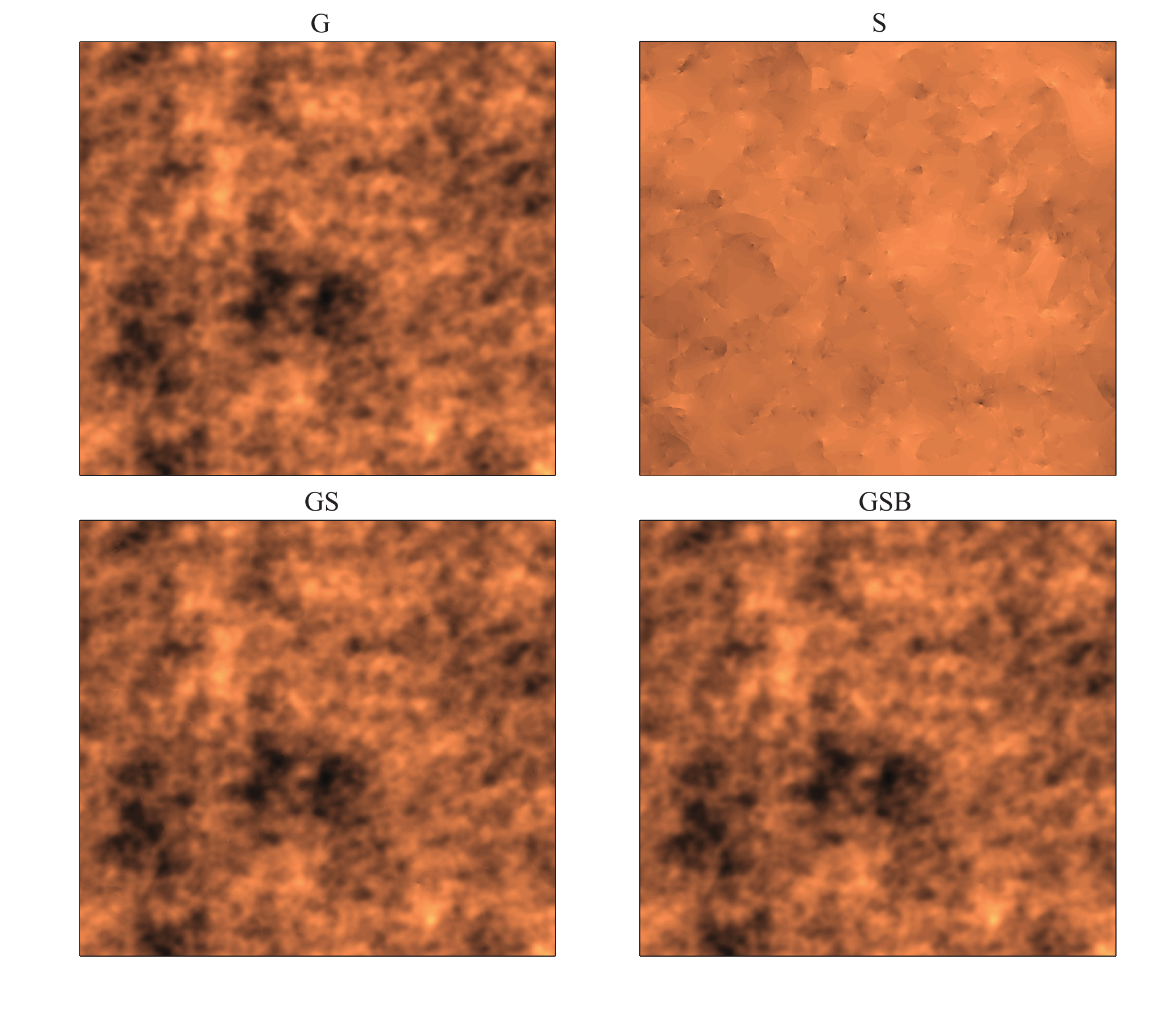}
\end{center}
\caption{Different components of our simulated maps. The map size is $7.2^{\circ}\times 7.2^{\circ}$ at resolution $R=0.42'$.
 The upper left plot is a  Gaussian CMB map, named by G in the text, simulated based on {\it Planck} 15 $\Lambda$CDM parameters. The upper right plot shows the CS-induced anisotropies, i.e., the S component,  with $G\mu=2.0\times 10^{-7} $. The lower left panel is the combination of the two, i.e., the GS map, smeared by the beam (Section~\ref{sec:beam}) in the lower right panel, making the GSB map.}
\label{fig:map2e7}
\end{figure}

 Figure~\ref{fig:power} compares the fiducial power spectrum as produced by CAMB (the solid black line) with the measured power spectrum from the simulated Gaussian maps (the filled circle symbols).
  Figure~\ref{fig:map2e7} illustrates various contributions to the simulations and their combinations.



\subsection{Cosmic string simulation}\label{sec:cs}
For the CS-induced CMB anisotropies, we use 100
high-resolution flat-sky CMB maps identical to the ones discussed by
\citep{Fraisse:2007nu}. They are obtained from numerical simulations of
Nambu-Goto string networks using the Bennett-Bouchet-Ringeval
code~\citep{Bennett:1990, Ringeval:2005kr} together with a direct
computation of the ISW effect generated by each string along the line
of sight. Unlike other numerical methods which are restricted to a short
redshift span, typically $\Delta z \approx 10^2$, and are thus only
reliable on large angular scales, these simulations are produced by
stacking maps from various redshifts  (outlined in
\citealt{Bouchet:1988hh, Ringeval:2012tk}), a valid approach for small
scale simulations. Among the main simplifying assumptions used in
these simulations is the small-angle approximation used in the
computation of the ISW effect from CSs~\citep{Stebbins:1988, Hindmarsh:1993pu, Stebbins:1995}.

The CS-induced anisotropies with the desired amplitude and the
inflationary Gaussian anisotropies are then combined to form our CMB
sky, without yet the instrumental effects being taken into
account. The CS tensions used in this work are in the range
$2.6\times10^{-11} \leq G\mu \leq 5.0\times10^{-7}$.  One can therefore
ignore the effect of string contribution on the CMB power spectrum in
the scales of interest in our analysis, without losing much precision.
This can be seen from Figure \ref{fig:power}. The long-dashed line
represents the power spectrum $C^{\uS}_{\ell}$ of CS contribution to
the fluctuations, expected to behave as
$\ell(\ell+1)C^{\uS}_{\ell}\sim \ell^{-\varepsilon}$ with $\varepsilon=0.90\pm0.05$ for
$\ell\gg 1$ \citep{Fraisse:2007nu}. A map of CMB anisotropies
generated by a CS network with $G\mu= 2.0\times 10^{-7}$ is
illustrated in the upper right panel of Figure \ref{fig:map2e7},
compared to the GS-realization (lower left panel). Careful visual
investigation of the plots reveals noticeable sharp edges from string
anisotropies.

\subsection{Instrumental noise}\label{sec:Gnoise}
Our model of the instrumental noise is a white Gaussian random field
characterized by the signal-to-noise ratio $SNR$, taken to be $10$,
$15$ and $20$. These noise levels are close to the instrumental noise
of the Atacama Cosmology Telescope (ACT), CMB-S4 phase I and CMB-S4 phase II,
respectively. The goal here is to see the overall impact of noise
contamination on the CS detectability. It is obvious that making
realistics forecast of the capability of any experiment in the search for
the CSs would require more realistic noise modeling.

\subsection{Beam}\label{sec:beam}
Due to the finite resolution of the telescopes, observed CMB temperature anisotropies are the result of the convolution of  underlying temperature distribution on the sky with the instrumental beam.
  In the following, we
  consider two types of experiment. The first is a {\it Planck}-like
  experiment in which the beam is modeled as a Gaussian with
  $\mathrm{FWHM}=5'$. This is the beam used for the {\it Planck}-like observational setup in this work. Based on the specification of the Millimeter Biometric Array Camera (MBAC) generation of CMB detectors used in ACT, a second type of beam is used for any other experimental strategy (i.e.,  the ACT-like, the CMB-S4 phase I- and II-like and the ideal, noise-free experiments).  We select the effective band
center 274 GHz with $\mathrm{FWHM}=0.9'$.

Following the notation of \citep{Fraisse:2007nu}, the Fourier  components of the observed CMB map, ${\cal V}(k)$, will be described by
\begin{equation}
{\cal V}(k)=\frac{\partial B_{\nu}(T)}{\partial T} T_{\rm CMB}\int\delta_T(r)A(r)e^{-i\mathbf{k}.\mathbf{r}} {\rm d}^2 r.
\end{equation}
Here, $k$ is the wave-number and $\mathbf{r}$ is the coordinate of a point on the telescope. Also $\delta_T(r)$, $B_{\nu}(T)$ and $A(r)$ represent respectively the CMB temperature anisotropy on the sky, the {\it Planck} function and the primary beam function, here taken to have an Airy pattern.
In the simulations, we use the Fourier transform of the beam per unit area, $\tilde{A}(u)$,  related to the beam itself through
\begin{equation}
A(r)=\frac{1}{(2\pi)^2}\int \tilde{A}(u) e^{2\pi
  i(\mathbf{u}.\mathbf{r})} \ud^2 u.
\end{equation}
with
\be\label{eq:beam:fourier}
\tilde{A}(u)=\mathcal{A}\left[\arccos\frac{u}{u_\uc}-\frac{u}{u_\uc}\sqrt{1-\left(\frac{u}{u_\uc}\right)^2}\right],
\ee
and $\mathcal{A}=2/(\pi^4d^2)$, $u\equiv\frac{k}{2\pi}$. The diameter of the telescope, $d$, is taken to be 6~{\rm m}. If the characteristic maximum opening of the telescope is $\theta$ (set to be $70^\circ$), then $u_\uc=\theta/\lambda$ with $\lambda$ being the wavelength of the observation. Eq.~\ref{eq:beam:fourier} defines  $\tilde{A}(u)$ only for $u\le u_\uc$ and $\tilde{A}(u)$ is zero elsewhere. The normalization guarantees that  $A(0)=1$.
In the small-scale regime, the multipole representation of Eq.~\ref{eq:beam:fourier} is
\begin{equation}\label{eq:beam1}
\tilde{A}(\ell)=\mathcal{A}\left[\arccos\frac{\ell}{\ell_\uc}-\frac{\ell}{\ell_\uc}\sqrt{1-\left(\frac{\ell}{\ell_\uc}\right)^2}\right],
\end{equation}
where $\ell_\uc=2\pi d/(\lambda \theta)$. The suppressing effect of the
beam on large multipoles is evident from Figure
\ref{fig:power}. The dotted line represents the beamed
power spectrum of CMB fluctuations whereas the dashed-dot-dot line
corresponds to the underlying temperature distribution on the sky.
This smearing effect is also recognizable on the map itself for small
scales (see Figure \ref{fig:map2e7}).

\section{Cosmic string detection pipeline}\label{sec:pipeline}

Our goal in this work is to evaluate the performance of various
sequences of image-processing and statistical tools in the detection
of the trace of a possible CS network on CMB temperature anisotropies.
We are interested in detecting line-like discontinuities in
temperature maps produced by the strings and curvelets are an
adequate tool for this purpose. Indeed, the basis functions of
curvelets are localized in both Fourier and position spaces. These
elongated basis functions enable curvelets to track well the CS footprints
on CMB maps \citep{fadili2009curvelets}.
The maps are then passed through a chain of filters to magnify their edge discontinuities.
Statistical measures and a P-value analysis are applied to these curvelet-decomposed and gradient maps to
assess the capability of the methods in detecting CSs contribution
for the various beam and noise levels associated with each experiment.
In brief, our proposed pipeline comprises two major steps:
\begin{itemize}
\item[1--] Processing CMB maps: here we apply several image-processors with
  the aim to isolate or/and enhance the CSs imprint on CMB maps (see
  Section~\ref{sec:processing}). The two pillars of this step are a
  multi-scaling analysis through curvelet-decomposition of the input
  maps (Section~\ref{sec:curvelet}) and the generation of filtered maps through extended Canny algorithm (ECA) (Section~\ref{sec:ECA}).
\item[2--] Analysis of processed CMB maps: here we use various statistical measures to quantify the detectability of CSs signature on the filtered maps from the first step
  (see Section~\ref{sec:stat_tools}).
\end{itemize}
The efficiency of the method can be summarized by the minimum detectable value of $G\mu$ for each sequence of steps.

\subsection{Processing CMB maps} \label{sec:processing}
In this section, we develop the image-processing part of our CS detection pipeline with the aim to increase the chances of the CS signal detectability.
\begin{figure}
\begin{center}
\includegraphics[scale=0.42]{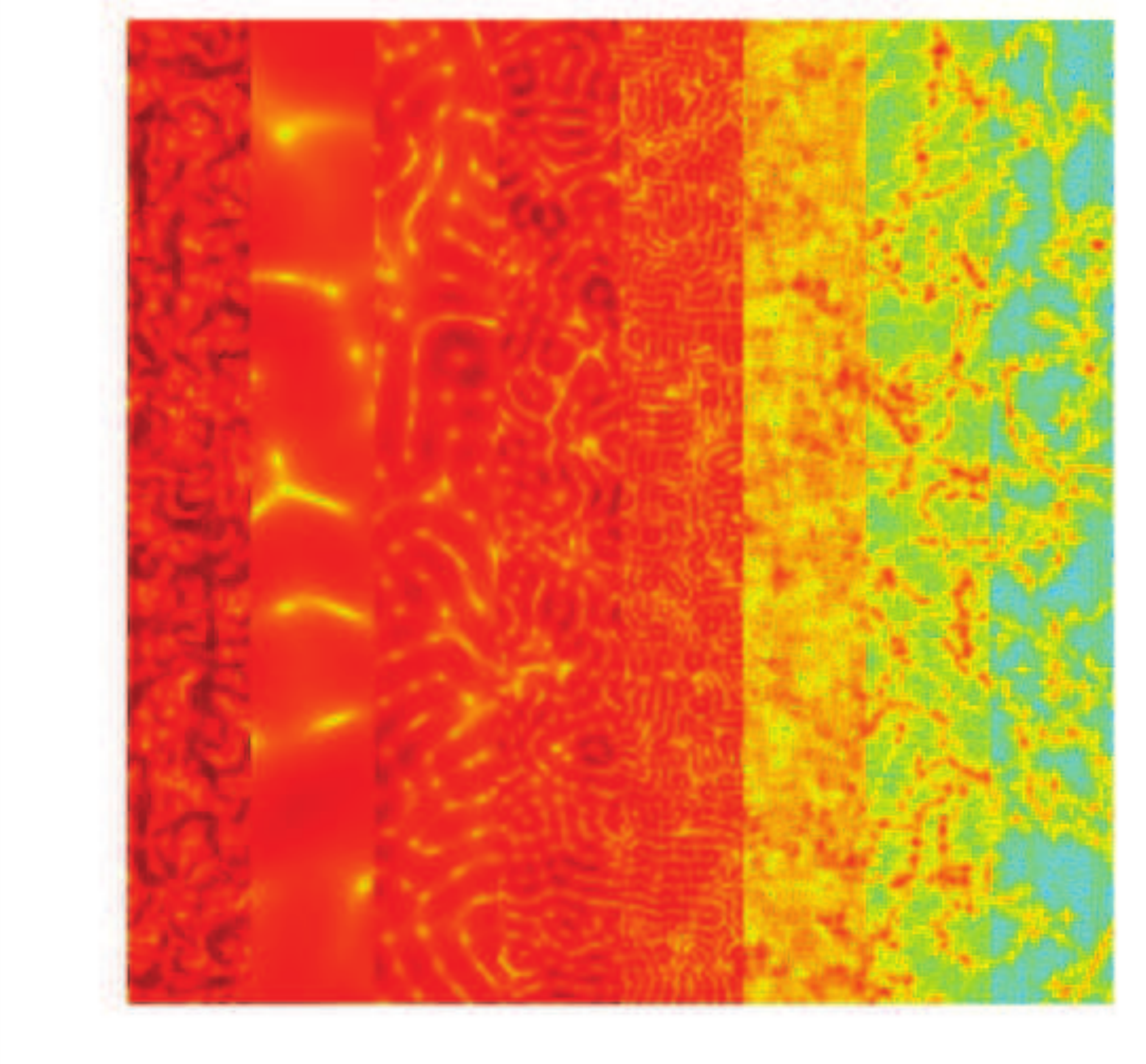}
\end{center}
\caption{Filtered GSB maps
  generated from different curvelet components of CMB
  anisotropies (beam of $0.9'$). From left to right, the first panel is the full map,
  and the rest are the first to the seventh curvelet components. The
  filter used for edge extraction is \textit{Scharr} \citep{jain1995machine}. 
    The CS network contribution to the fluctuations corresponds
  to $G\mu=1.0\times 10^{-7}$. The visual detectability of the
  string-induced discontinuities, especially in the panels with finer
  scales (i.e., in the higher curvelet modes) is striking.  The
  color scale is logarithmic. }
\label{fig:eightcomponent}
\end{figure}

\subsubsection{Multi-scaling analysis: curvelet decomposition of CMB maps}\label{sec:curvelet}

The expansion of a field or function in a complete set of basis functions has a long history in various fields of study, with the most familiar one known as the Fourier transformation.
The basis functions for Fourier transformation are maximally localized in wavenumber (frequency) space.
Therefore, the individual Fourier components have no information about local events in position (time) space.
To resolve this limitation, generalizations of Fourier transformations, such as wavelet and ridgelet transforms, have been developed.

  Wavelets are developed with localized basis functions  in both position (time) and wavenumber (frequency) spaces.
  They provide an excellent mathematical architecture for sparse representation of data with transient or local features, where too many Fourier modes would otherwise be required.
  Wavelets accomplish this by using  multi-scale, local basis functions which are isotropically extended. In order to assess anisotropic features embedded in the underlying field, additional modification are required.
    More specifically, the orientation selectivity of wavelet is weak and, in two or three dimensions, can not  efficiently represent curve-like singularities.
    For example, the basic 2D wavelet is not able to identify elongated features such as signature of the CSs as discontinuities along the edges.

  Alternative transformations, which are capable of overcoming this limitation, are ridgelets and curvelets, firstly introduced and developed by \citep{candes2000curvelets,donoho2000digital,candes2001curvelets}.
  They are relatively new in the wavelet family and are already widely used in various fields from medicine to physics (see \citep{fadili2009curvelets} for a comprehensive introduction).
Ridgelet transformations take into account scales and positions as well as orientations, and are ideal for straight line detection.

Curvelets, on the other hand, are commonly used when some sort of smooth curve detection is required.
Intuitively, the curvelet transform is a multi-scale pyramid with enough directions and positions at each given length scale, and needlet-shaped elements at fine scales.
Therefore, curvelets decompose two and three dimensional images and
data sets into contributions from different scales, locations and
directions. They are different from other directional wavelets, such
as countourlets and directionlets, in that the degree of directional
localization is scale-dependent. This characteristic makes curvelets
ideal for spare representation of images which are smooth except for
curve-like discontinuities or edges.
The first generation of curvelets (curveletG1), based on local ridgelet transforms, could extract edges in suppressed backgrounds \citep{donoho2000digital,candes2001curvelets}, with significant improvement in the next version (curveletG2)  where a mother (prototype) curvelet function is used for computing the expansion coefficients \citep{candes2004new,candes2002new}.
 Using unequally Spaced Fast Fourier Transform, \citep{candes2006fast} developed two fast discrete curvelet transforms which are simpler, faster and more efficient compared to other approaches, and have both discrete  and continuous versions

 The general curvelet transformation of a square integrable two dimensional map $T\in{\rm L}^2(\mathbb{R}^2)$ is given by:
 \begin{eqnarray}
 T(x,y)=\sum_{j,k,l}\langle T,\phi_{jkl}\rangle \phi_{jkl}
 \end{eqnarray}
 where the $\phi_{jkl}$'s are the curvelet basis functions. The curvelet coefficients, represented by $\langle T,\phi_{jkl}\rangle$ are given by the ${\rm L}^2$-scalar product of the map $T$ and the $\phi_{jkl}$'s.
  The three indices $j$, $k$ and $l$ represent scale, orientation  and location (in position space), respectively.
  In this work we use \textit{CurveLab}\footnote{Available at \url{http://www.curvelet.org/}. It contains the \textit{Matlab} and \texttt{C++} implementations of both the USFFT-based and the wrapping-based transforms.}, the 2D discrete version of the curvelet transform.
  We wrapped the original package in Python, called {\it Pycurvelet}, which is available upon request.

{\it CurveLab} applies the Fast Fourier Transform (FFT) on the data. The resulting 2D Fourier map is then divided into wedges through slicing by concentric circles and angular divisions. This procedure decomposes the map into multiple scales and different orientations. Each wedge, produced this way, corresponds to a certain curvelet component, associated with a particular scale and orientation. Applying inverse FFT on a given wedge leads to the corresponding curvelet coefficient for the associated scale and orientation at a given point.

Among the adjustable parameters used in the decomposition are the number of scales $n_{\rm scales}$ and orientations $n_{\rm angles}$. Higher values of these parameters correspond to more components and higher resolutions.  In the trade-off between the computational cost and the desired accuracy of the results, we found the appropriate parameters for our work to be  $n_{\rm scales}=7$ and $n_{\rm angles}=10$.
Figure \ref{fig:eightcomponent} illustrates the seven curvelet components of a simulated CMB sky with contribution from the CS network with $G\mu=1.0\times 10^{-7}$, compared to the full map itself (the leftmost bar). The trace of the CS network is visually distinguishable for components with high mode number.

\begin{figure}
\begin{center}
\noindent
\noindent\begin{minipage}[ht]{0.2\linewidth}
\begin{center}
\resizebox{\columnwidth}{!}{%
\begin{tabular}{ |c|c|c|c| }
\hline
0 & 0 & 0 \\
\hline
0 & 1 & -1 \\
\hline
0 & 0 & 0 \\
\hline
\end{tabular}
}\\
\smallskip
$D_x$
\end{center}
\end{minipage}
\quad
\noindent\begin{minipage}[ht]{0.2\linewidth}
\begin{center}
\resizebox{\columnwidth}{!}{%
\begin{tabular}{ |c|c|c|c| }
\hline
0 & 1 & 0 \\
\hline
1 & -4 & 1 \\
\hline
0 & 1 & 0 \\
\hline
\end{tabular}
}\\
\smallskip
$\mathcal{L}$
\end{center}
\end{minipage}
\quad
\noindent\begin{minipage}[ht]{0.2\linewidth}
\begin{center}
\resizebox{\columnwidth}{!}{%
\begin{tabular}{ |c|c|c|c| }
\hline
-1 & 0 & 1 \\
\hline
-2 & 0 & 2 \\
\hline
-1 & 0 & 1 \\
\hline
\end{tabular}
}\\
\smallskip
$G_x$
\end{center}
\end{minipage}
\quad
\noindent\begin{minipage}[ht]{0.23\linewidth}
\begin{center}
\resizebox{\columnwidth}{!}{%
\begin{tabular}{ |c|c|c|c| }
\hline
3 & 0 & -3 \\
\hline
10 & 0 & -10 \\
\hline
3 & 0 & -3 \\
\hline
\end{tabular}
}\\
\smallskip
$H_x$
\end{center}
\end{minipage}

\end{center}
\caption{Different filters for edge recovery applied on the anisotropy maps after curvelet decomposition. Left to right:  the standard unit neighborhood derivative in the $x$ direction, the {\it Laplacian} operator,  the {\it Sobel} operator in the $x$ direction and the {\it Scharr} operator in $x$ direction. The "$y-axis$" operators are made similar to "$x-axis$" with proper rotations.}
\label{fig:filters}
\end{figure}

\begin{figure*}
\begin{center}
\includegraphics[scale=0.25]{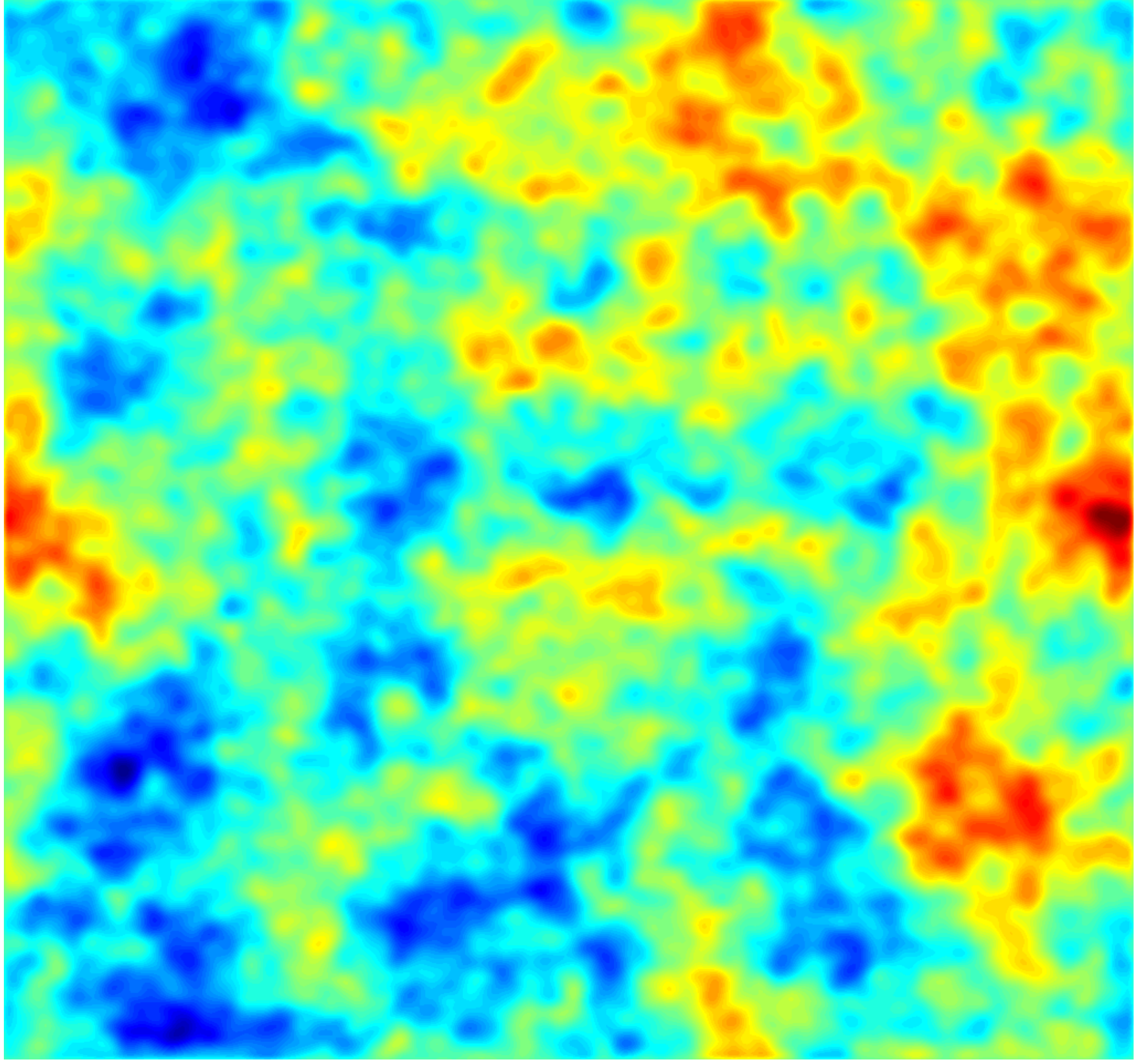}
\includegraphics[scale=0.25]{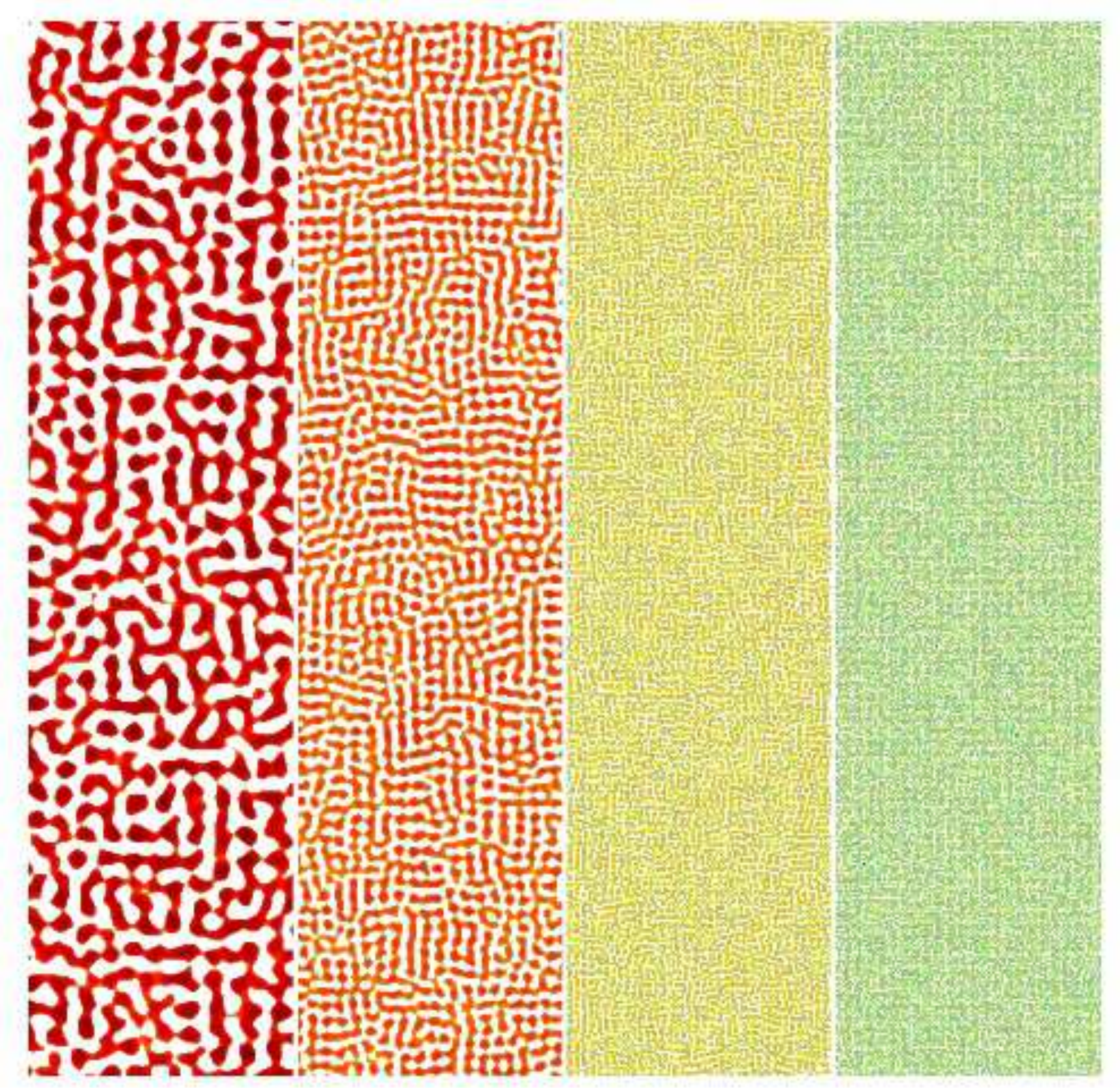}
\includegraphics[scale=0.25]{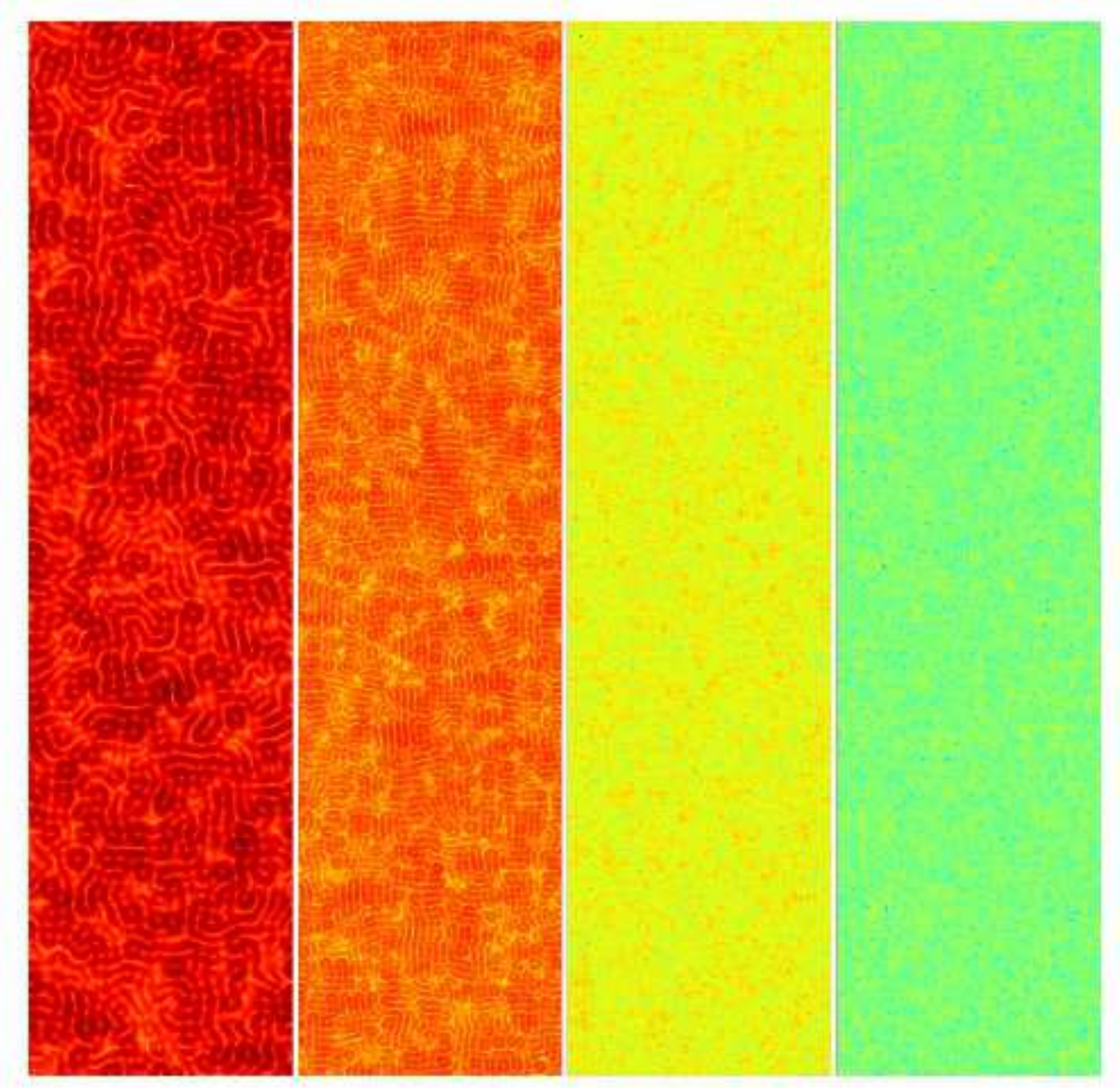}

\includegraphics[scale=0.25]{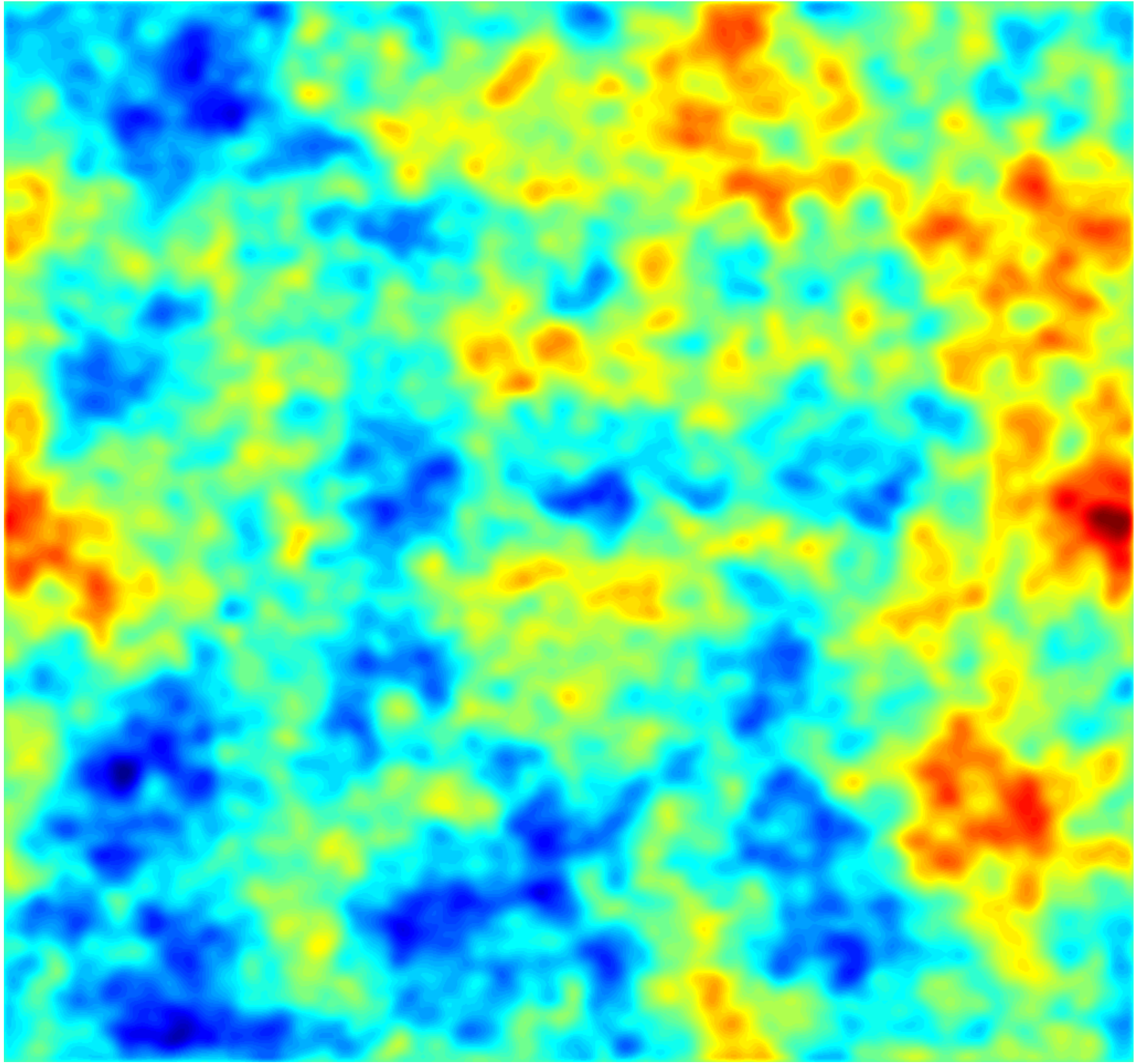}
\includegraphics[scale=0.25]{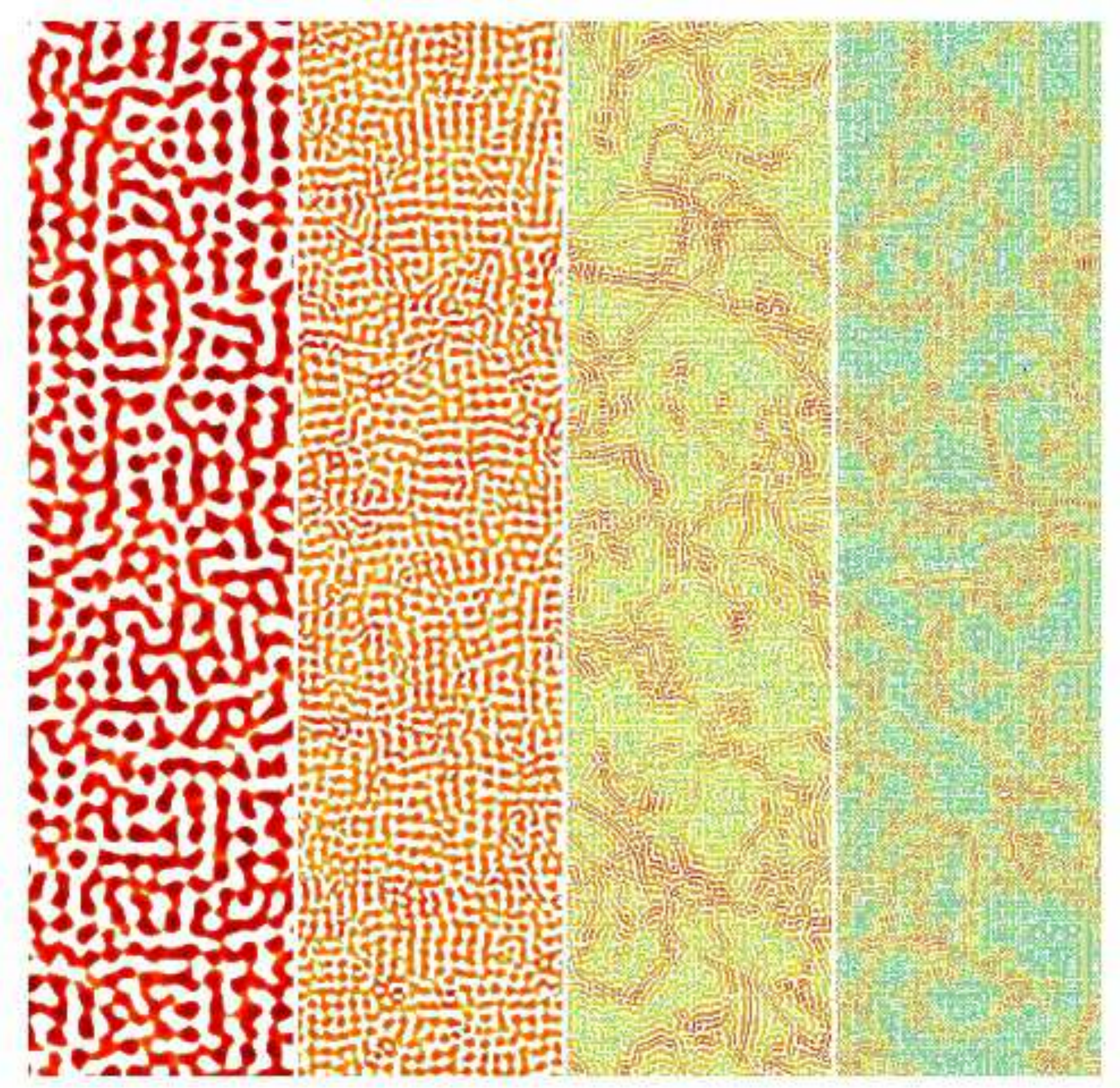}
\includegraphics[scale=0.25]{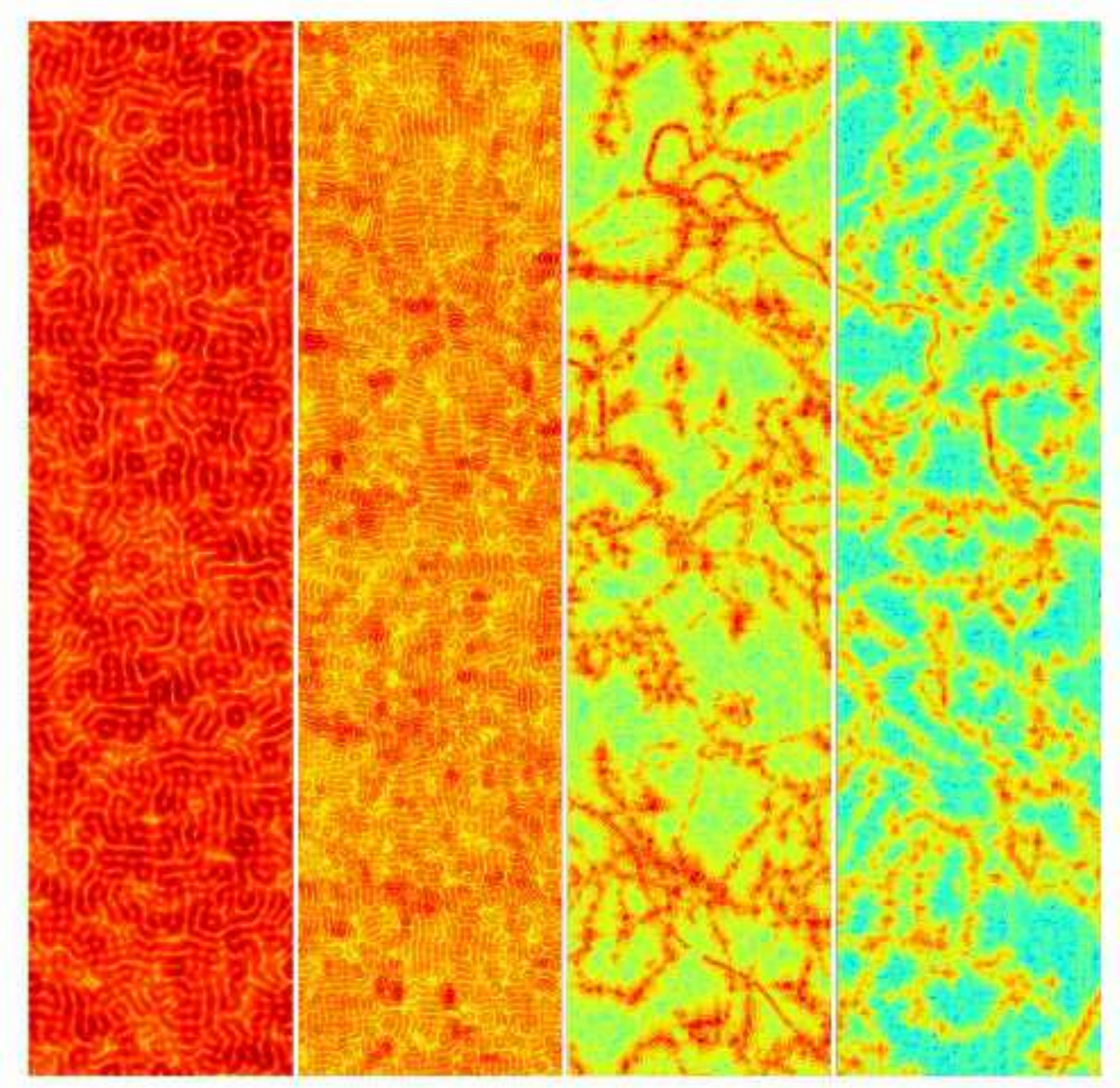}

\includegraphics[scale=0.25]{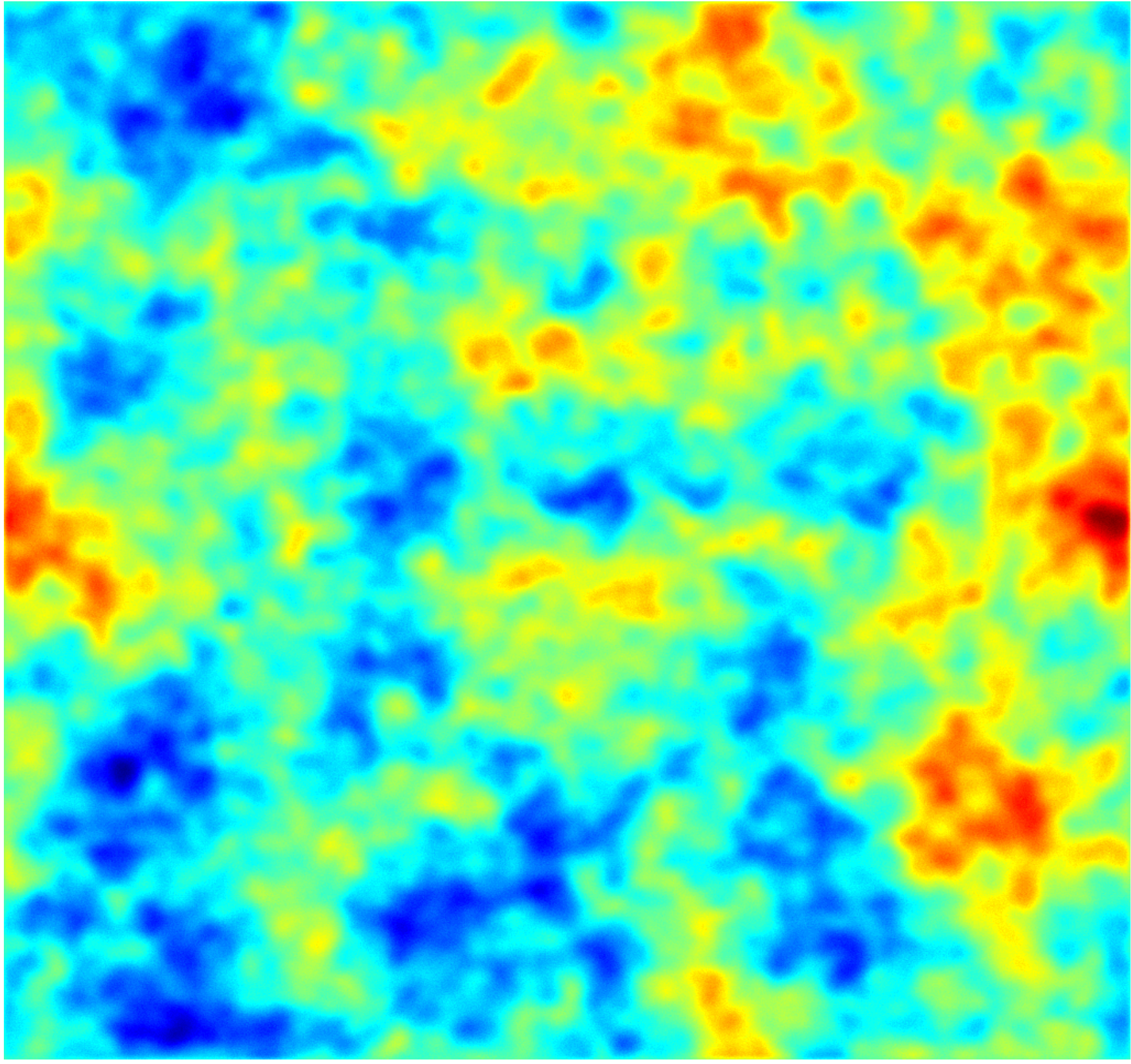}
\includegraphics[scale=0.25]{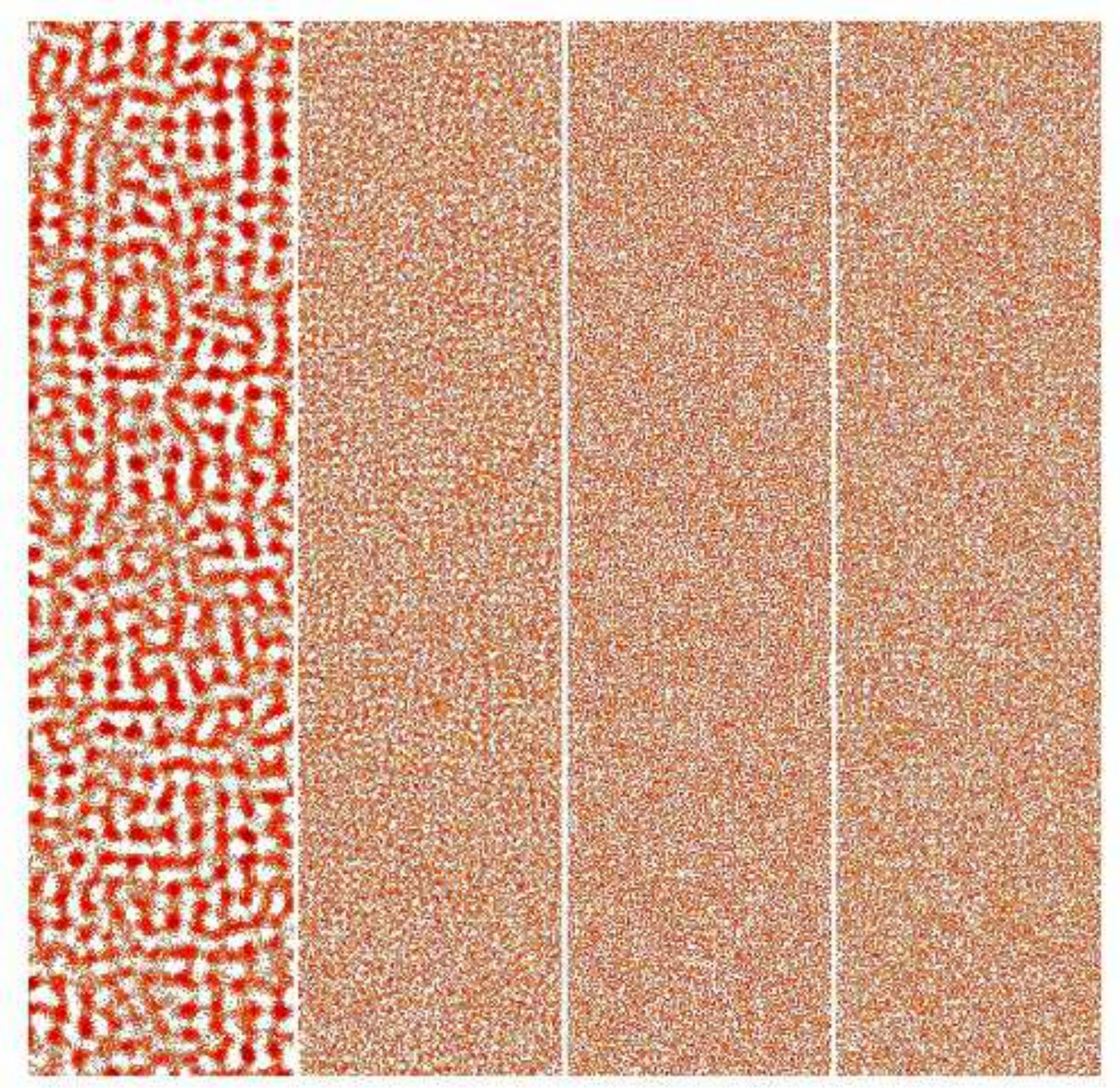}
\includegraphics[scale=0.25]{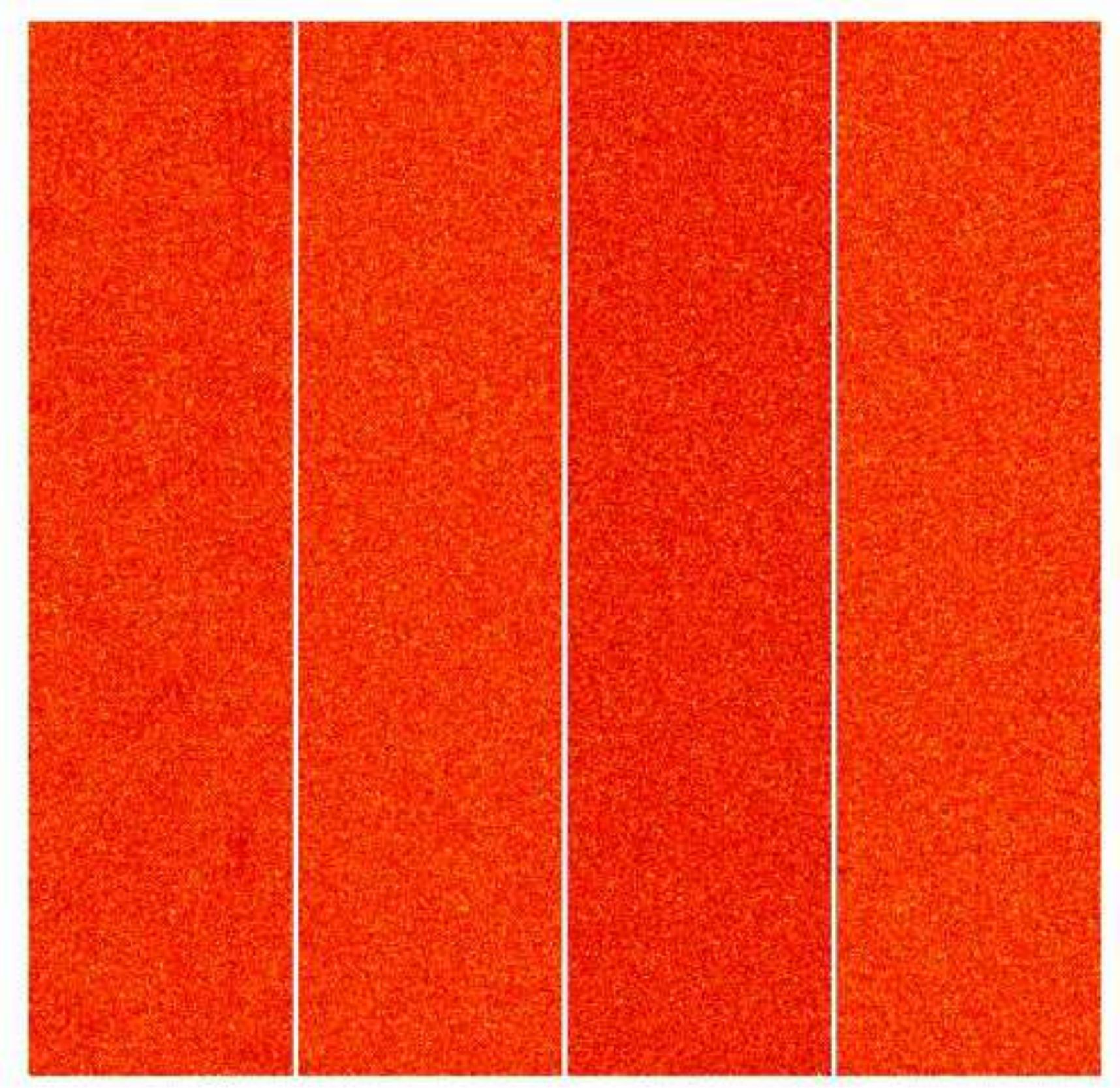}
\end{center}
\caption{Visual evaluation of the image-processing steps of the proposed CS-detection pipeline. The top, middle and bottom rows correspond respectively to GB maps, GSB maps with $G\mu=1.0\times 10^{-7}$, and GSBN maps with $G\mu=1.0\times 10^{-7}$ and $SNR=20$  (similar to the noise level of a CMB-S4 phase II experiment).
From left to right, the panels are the full maps, the fourth to the seventh curvelet components and their edge maps produced by  {\it Scharr} filter. The CS trace, not visually distinguishable in the map, is clearly detectable in the last two components, with significantly boosted detectability after filtering. }
\label{fig:edge_maps}
\end{figure*}

\subsubsection{Filtering the maps: extended Canny algorithm}\label{sec:ECA}

The imprint of the CS network on CMB anisotropies can be thought of as (the superposition of) line-like structures, conveniently characterized by sharp discontinuities known as  edges.
Among the widely used algorithms developed to identify edges in 2D images is the Canny edge detector \citep{canny1986computational} which is a multi-stage method.
 The image is initially smoothed by a Gaussian kernel to reduce the intrinsic noisiness due to the stochastic nature of the field.
 The edges are defined as  points with large gradients. Therefore, the central piece of the algorithm is to find the image gradient.

The edge-identification step of our CS-detection pipeline is based on the Canny algorithm, but is extended in certain ways, and is therefore called  the extended Canny algorithm (ECA).
The extensions include using kernels other than the Gaussian to
smooth the maps, including Boxcar and Hanning \citep{blackman1958measurement}. 
We also apply various filters, including {\it Derivative} (der), {\it Laplacian} (lap), {\it Sobel} (sob) \citep{jahne1999handbook} and {\it Schaar} (sch) \citep{jain1995machine}, to produce gradient maps. 
Figure \ref{fig:filters} illustrates how these  filters act on the neighboring pixels to construct the gradient map.

Our ultimate goal is to perform the ECA on curvelet components of the CMB anisotropy maps and evaluate their performance in identifying the CS-induced edge-like features.
Figure \ref{fig:edge_maps} neatly demonstrates how these two successive steps isolate/enhance the sought-after CS signature on the original CMB maps, and thus prepare it to undergo the next step, i.e., statistical analysis.
The rows in Figure \ref{fig:edge_maps} represent simulations of  GB,  GSB (with $G\mu=1.0\times 10^{-7}$) and GSBN (with $G\mu=1.0\times 10^{-7}$ and $SNR=20$) cases, from top to bottom, respectively.
The columns on the other hand, illustrate the image processing steps: the leftmost panel shows the input maps, the middle represents their (forth to seventh) curvelet components, and the rightmost corresponds to the gradient maps produced from these components, using {\it Scharr} as the ECA filter.
The footprints of the CS network are evident in the curvelet components and more vividly in the filtered maps. Adding noise smears out these footprints, making them no longer visually distinguishable.
This necessitates developing proper statistical tools for their detection, as we do in the next section. We will also assess the sensitivity of our detection procedure to various filters for different input maps in Section~\ref{sec:results}.

\subsection{Analysis of the processed CMB maps} \label{sec:stat_tools}
Applying the preprocessing steps of Section~\ref{sec:processing} on
CMB maps increases the detectability of their possible CS signature,
first, through keeping only components with the largest contribution
from strings, and secondly, by locating the edges, assumed to be
induced by the CS network.
Equipped with filtered CMB temperature maps, we now turn to the last step of our pipeline, i.e., measuring certain statistical properties of the maps to quantify the detectability of their CS imprint.

\subsubsection{Notation}\label{sec:stat} \label{sec:stat}

Here we outline some definitions and introduce our notation, used in the rest of the paper.
The CMB temperature anisotropies is a stochastic field,
represented by a 2D map $T\in{\rm
    L}^2(\mathbb{R}^2)$ which is obtained according to Eq.~(\ref{fullmap}).
One can construct a vector $\mathcal{A}$ at each spatial point as:
$$\{\mathcal{A}\}\equiv\{\delta_T,\eta_{x},\eta_{y},\xi_{xx},\xi_{yy},\xi_{xy}\}$$
where $\delta_T$ is the {\it density contrast} of the stochastic
field. For the CMB anisotropies here $\delta_T\equiv T$ (temperature fluctuation).
We have also defined  $\eta_{x}\equiv \partial \delta_T/\partial x$, $\eta_{y}\equiv \partial \delta_T/\partial y$ and $\xi_{xy}\equiv\partial^2 \delta_T/\partial x\partial y$.  In general, ${\mathcal A} $ can be expanded to include higher order derivatives. On the other hand, in certain cases, the first order derivative may suffice to explore the statistical feature one is interested in. For example, studying the crossing statistics only requires the knowledge of the first order derivatives while peak analysis requires the second order as well.

The characteristic function of  $\mathcal{A} $, intimately related to its free energy, is defined by:
\begin{eqnarray}
\mathcal{Z}(\lambda)=\int_{-\infty}^{+\infty}\ud^6{\mathcal{A}
  {\mathcal{P}}({\bf{\mathcal{A}}}}) \bfmath{e}^{{\bfmath{i}}\lambda.{\bf{\mathcal{A}}}},
\end{eqnarray}
where $\lambda$ is an array with the same dimension as $\mathcal{A}$.
$\mathcal{Z}$ can be expanded as \citep{Matsubara:2003yt}
\begin{eqnarray}\label{parti1}
&&\mathcal{Z}(\lambda)=\exp{ \left( {-\frac{1}{2}\lambda^T.\mathcal{C}.\lambda}\right)}\nonumber\\
 &\times& \exp{\left[ \sum_{j=3}^{\infty}\frac{{\bfmath{i}}^j}{j!}\left(\sum_{\mu_1}^N\sum_{\mu_2}^N...\sum_{\mu_j}^N \mathcal{K}^{(j)}_{\mu_1,\mu_2,...,\mu_j}\lambda_{\mu_1}\lambda_{\mu_2}...\lambda_{\mu_j}\right) \right] }. \nonumber\\
\end{eqnarray}
where ${\mathcal K}^{(n)}_{\mu_1,\mu_2,...,\mu_n}\equiv\langle \mathcal{A}_{\mu_1}\mathcal{A}_{\mu_2}...\mathcal{A}_{\mu_n} \rangle_\uc$  are the
cumulants and  $\langle \rangle_\uc$ stresses that only {\it connected} cumulants are taken into account.
Here $N$ is the dimension of $\mathcal{A}$ and throughout this paper $N=6$. Also $\mathcal{C}\equiv \langle {\mathcal A} \otimes{\mathcal A}\rangle$ represents the $6\times6$ covariance matrix of ${\mathcal A}$ at each spatial point.
Note that with the zero-mean CMB fluctuations the cumulants are the same as moments.
Various spectral parameters for the CMB field are defined
by
\begin{eqnarray}\label{spec2}
\sigma_m^2&\equiv& \langle \nabla^m\delta_T \nabla^m\delta_T \rangle\nonumber\\
&=& \frac{1}{(2\pi)^2} \int \ud{\bfmath{k}}k^{2m}P_{TT}(k)\tilde{W}^2(kL),
\end{eqnarray}
for small sky coverage. In this expression, $\tilde{W}$ stands
  for any smoothing function, such as the beam, and $L$ is the smoothing scale.

\subsubsection{Statistical measures}\label{sec:stats}
Here we introduce the statistical tools used in this work to quantify
the detectability of the imprints left by the CS network on CMB
anisotropies.

\begin{itemize}
\item[1--]\textit{The one-point PDF}

The one-point probability density function (hereafter, the PDF) of a
distribution describes the statistical abundance of the field values
and can be calculated from the inverse Fourier transform of the
characteristic function. For the joint probability density function
(JPDF) of $\mathcal{A}$ we have
\begin{eqnarray}
  &&\mathcal{P}({\bf{\mathcal{A}}})=\frac{1}{(2\pi)^6}\int_{-\infty}^{+\infty}
  \ud^6\lambda
  \mathcal{Z}(\lambda){\bfmath{e}}^{-{\bfmath{i}}\lambda.{\bf{\mathcal{A}}}}.
\label{parti2}
\end{eqnarray}
%
Plugging Eq.~\eqref{parti1} in Eq.~\eqref{parti2} gives
\begin{equation}
\begin{aligned}
  \mathcal{P}({\mathcal{A}})=
  \exp \bigg[\sum_{j=3}^{\infty}\frac{(-1)^j}{j!}
      \bigg(\sum_{\mu_1=1}^6 &...\sum_{\mu_j=1}^6 {\mathcal
        K}^{(j)}_{\mu_1,\mu_2,...,\mu_j} \\ & \times \frac{\partial^j}{\partial \mathcal{A}_{\mu_1}...\partial \mathcal{A}_{\mu_j}}\bigg)\bigg] {\mathcal {P}}_{\rm G}({\mathcal{A}})\,,
\end{aligned}
\end{equation}
where
\begin{equation}
 {\mathcal {P}}_{\rm G}({\mathcal{A}})=\frac{1}{\sqrt{(2\pi)^{6} |\mathcal{C}|}} \
{\bfmath{e}}^{-\frac{1}{2}({\bf \mathcal{A}}^{T}.\mathcal{C}^{-1}.{\bf
    \mathcal{A}})}.
\end{equation}
 The anisotropies produced by the CS network are
 non-Gaussian~\citep{Ringeval:2010ca}. The perturbative form of the one-point
 PDF of the temperature fluctuations,
 $\mathcal{P}_{\delta_T}(\alpha)$, in the presence of CSs, keeping
 only terms up to $\mathcal {O}(\sigma_0^3)$, is given
 by
\begin{equation}
\begin{aligned}
  {\mathcal{P}}_{\delta_T}(\alpha)&=\left \langle
  \delta_\ud(\delta_T-\alpha)\right \rangle = \int \ud^6 \mathcal{A}
  \delta_\ud(\delta_T-\alpha) \mathcal{P} (\mathcal{A}) \\
&= \frac{1}{\sqrt{2\pi}\sigma_0}{\bfmath{e}}^{-\alpha^2/2\sigma_0^2}\left[1+A\sigma_0+B\sigma_0^2+\mathcal{O}(\sigma_0^3)\right].
\label{pdf1}
\end{aligned}
\end{equation}
Here $A\equiv \frac{S_0}{6}H_3\left(\frac{\alpha}{\sigma_0}\right)$ and $B\equiv \frac{K_0}{24}H_4\left(\frac{\alpha}{\sigma_0}\right)+\frac{S_0^2}{72}H_6\left(\frac{\alpha}{\sigma_0}\right)$. Also $S_0\equiv\langle\delta_T^3\rangle_\uc /\sigma_0^4$ and $K_0\equiv\langle\delta_T^4\rangle_\uc /\sigma_0^6$ are the modified skewness and kurtosis quantities, respectively. The $H_n(\delta_T/\sigma_0)$ represents the probabilistic's Hermite polynomial of order $n$.
Note that the $G\mu$ levels we are interested in have tiny contributions to the CMB fluctuations compared to inflation-induced anisotropies. However, we show that proper sequences of image-processing and statistical steps can explore these regimes and possibly detect the tiny imprints.
\item[2--]\textit{The (weighted) TPCF}

The (weighted) two-point correlation function (TPCF) is defined as
 ${\mathcal C}_{TT}=\langle \delta_T(\textbf{r}_1)\delta_T(\textbf{r}_2)\rangle $
where $\textbf{r}_1$ and $\textbf{r}_2$ represent the coordinates of
the points. ${\mathcal C}_{TT}$ is another statistical measure we employ in this
work to  search for possible deviation from the ${\mathcal C}_{TT}$ produced by inflationary anisotropies.
\item[3--]\textit{The unweighted TPCF of peaks}

 Topological and geometrical criteria to characterize morphology of cosmological stochastic fields in one, two and three dimensions, have been considered in various researches (see e.g. \citealt{Matsubara:2003yt,Ducout:2012it,Pogosyan:2008jb, Gay:2011wz,Codis:2013exa}). The clustering of these  measures based on their TPCF also provides a useful statistical framework. Here we focus on the local maxima clustering.
The unweighted  TPCF of a certain feature of a stochastic field, also referred to as its excess probability, is a robust measure of the clustering of that feature.
From the statistical-mechanics point of view, the information about an interacting system is encoded in the excess probability of finding certain features of the stochastic field representing that system.
In this paper we compare the clustering of the local maxima of CMB maps for Gaussian-only fluctuations with those including contributions from the CS network as well.
The excess probability of finding peak pairs $\Psi_{\rm pk-pk}$ separated by distance $r=|\textbf{r}_1-\textbf{r}_2|$, at thresholds $\vartheta_1\equiv \alpha_1/\sigma_0$ and $\vartheta_2\equiv \alpha_2/\sigma_0$ is defined as
\begin{equation}
\Psi_{\rm pk-pk}(r;\vartheta_1,\vartheta_2)=\frac{\left \langle n_{\rm
    pk}(\textbf{r}_1,\vartheta_1) n_{\rm pk}(\textbf{r}_2,\vartheta_2)
  \right \rangle}{\bar{n}_{\rm pk}(\vartheta_1)\bar{n}_{\rm pk}(\vartheta_2)} -1,
\end{equation}
where $\bar{n}_{\rm pk}({\vartheta})$ is the number density of peaks and is mathematically  given by
\begin{equation}
\bar{n}_{\rm pk}(\vartheta)=\left\langle \delta_\ud (\delta_T-\vartheta \sigma_0)\delta_\ud(\eta)\left|\det(\xi)\right|\right\rangle,
\end{equation}
The second derivative tensor of the CMB field ($\xi_{ij})$ should be {\it negative definite} at peak position. Its analytical expression for a 2D homogenous Gaussian field was calculated in \citep{Bardeen:1985tr,Bond:1987ub}.
An estimator for this excess probability, $\tilde{\Psi}_{\rm pk-pk}(r;\vartheta)$, is given by
\begin{eqnarray} \label{eq:pp-estimator1}
\tilde{\Psi}_{\rm pk-pk}(r;\vartheta) = \left[\dfrac{DD(r,\vartheta)}{RR(r,\vartheta)}
  \right] \dfrac{N_R(N_R-1)}{N_D (N_D -1)} -1,
\end{eqnarray}
which usually reduces  the boundary effect
 \citep{Landy:1993yu}. Here, $RR(r,\vartheta)$ and $DD(r,\vartheta)$ are the number of peak pairs in random and data catalogs, respectively, separated by distance $r$ from each other.  Similarly, $N_D$ and $N_R$ are the total number of peaks in data and random catalogs, respectively.
 %
\item[4--]\textit{The unweighted TPCF of up-crossings}

The crossing statistics was first introduced by \citep{rice1944mathematical}. Since then, it has been used to study
the geometry of stochastic fields in various disciplines, e.g., in
complex systems
\citep{brill2000brief,peppin1994introduction,jafari2006level,vahabi2011analysis},
material sciences \citep{nezhadhaghighi2015crossing},  optics
\citep{goodman2007speckle,yura2010mean,pirlar2017crossing} and
cosmology and early
universe~\citep{Ryden:1988rk,ryden1988collapse,Matsubara:1995wj,Movahed:2010zq,Matsubara:2003yt,
  Musso:2013pha, Musso:2014jda}.
Crossing statistics can be introduced for 1, 2 and 3D stochastic fields. For 1D it corresponds to crossing at a given threshold. Length or contour statistics corresponds to crossing statistics  for a typical 2D field, while for 3D, area statistics is representative of crossing statistics.  To be more specific, to up-cross a given threshold refers to when the field at a point crosses the threshold with a positive slope  (in a certain direction in a 2D field). In an isotropic stochastic field, up-crossing and down-crossing (i.e., crossing with a negative slope) are statistically equivalent.
The probabilistic framework of the mean number of up-crossings of a 2D field $\delta_T$  at a threshold  $\vartheta$  for an arbitrary 1D slice denoted by $\otimes$ is given by:
\begin{eqnarray}\label{theory for nu nongaussain}
{\bar n}_{\rm up}^{\otimes}(\vartheta)=\left\langle \delta_\ud(\delta_T-\vartheta\sigma_0)\Theta(\eta^{\otimes})\left|\eta^{\otimes}\right|
\right \rangle,
\end{eqnarray}
where $\Theta(\eta)$ is the unit step function and $|\eta^{\otimes}|$ is the absolute value of the first derivative of temperature fluctuations in direction $\otimes$ (e.g., see \citealt{Matsubara:2003yt}). For statistically isotropic CMB map, one can choose any direction $\otimes$ on the map,  and work in that direction with the one dimensional notion of the up-crossing, without loss of generality.
 For a pure Gaussian CMB  stochastic field, we have
\begin{eqnarray}\label{theory for nu gaussain}
\bar{n}^{\otimes}_{\rm up}(\vartheta)=\frac{1}{2\pi\sqrt{2}}\frac{\sigma_1}{\sigma_0}{\bfmath{e}}^{-\vartheta^2/2},
\end{eqnarray}
where $\sigma_0$ and $\sigma_1$ are spectral parameters defined by Eq. (\ref{spec2}).
 In this work, we go beyond the one-point statistics of up-crossings, $\bar{n}^{\otimes}_{\rm up}(\vartheta)$, and
 investigate their clustering as well, characterized by the excess probability of finding a pair of up-crossings separated by distance $r=|\textbf{r}_1-\textbf{r}_2|$, at thresholds $\vartheta_1$ and $\vartheta_2$
\begin{equation}
\Psi_{\rm up-up}(r;\vartheta_1,\vartheta_2)=\dfrac{\left \langle
  n_{\rm up}(\textbf{r}_1,\vartheta_1) n_{\rm
    up}(\textbf{r}_2,\vartheta_2) \right \rangle}{\bar{n}_{\rm up}(\vartheta_1) \bar{n}_{\rm up}(\vartheta_2)}-1,
\end{equation}
where $\bar{n}_{\rm up}(\vartheta)=\langle \bar{n}_{\rm up}^{\otimes}(\vartheta)\rangle $ and the averaging is over all available directions.
\\
\item[5--]\textit{The unweighted cross-correlation of up-crossings and peaks}

 We define the cross-correlation of peaks and up-crossings as:
\begin{equation}
\Psi_{\rm up-pk}(r;\vartheta_1,\vartheta_2)=\dfrac{\left \langle
  n_{\rm up}(\textbf{r}_1,\vartheta_1) n_{\rm
    pk}(\textbf{r}_2,\vartheta_2) \right \rangle}{\bar{n}_{\rm up}(\vartheta_1)\bar{n}_{\rm pk}(\vartheta_2)}-1.
\end{equation}
\\

\end{itemize}

\subsubsection{Quantifying the CS-induced deviation}

In the following, we apply the five statistical measures introduced above on the ECA gradient maps. These are generated through applying various filters on different curvelet components of simulated CMB maps.
In order to investigate the detectability of  the (enhanced) CS footprints on these gradient maps,
we define the following $\Delta_{\diamond}$'s. This quantity enables
us to quantify the deviation from pure-inflationary simulations of a
certain measure (labeled by $\diamond$) calculated for maps with
different $G\mu$'s.
 To avoid bias in $\Delta_{\diamond}$, ${\mathcal C}_{TT}(r)$, $\Psi_{\diamond}(r,\vartheta;\times)$ and $\mathcal{P}(\alpha)$ are generated by averaging over  1000 null cases, i.e., with $G\mu=0$,  simulations.

For the PDF and the weighted TPCF, we define
\begin{equation}
\begin{aligned}
  \Delta_{\rm PDF}^{\times}(G\mu) & \equiv \int \ud \alpha \left|\mathcal{P}(\alpha;\times;G\mu)-\mathcal{P}(\alpha;\times;G\mu=0)\right|,\\
\Delta_{\rm TT}^{\times}(G\mu) &\equiv  \int \ud r \left|C_{TT}(r;\times;G\mu)-C_{TT}(r;\times;G\mu=0)\right|.
\end{aligned}
\end{equation}
The symbol $\times$ represents the specific sequence of curvelet component and ECA filter being used. Also $\alpha\equiv\vartheta \sigma_0$.
For the TPCF we define
\begin{equation}
\Delta_{\diamond}^{\times}(G\mu)\equiv\sum_{\vartheta}\int \ud r \left|\Psi_{\diamond}(r,\vartheta;\times;G\mu)-\Psi_{\diamond}(r,\vartheta;\times;G\mu=0)\right|.
\end{equation}
Here "$\diamond$" can be  "pk-pk", "up-up" or "up-pk".

Finally we can compare simulations with string-induced fluctuations  with null sets.  In the next section we will present our results by reporting the minimum detectable value of $G\mu$  in CMB observations for various noise levels, using different parameter settings in the proposed pipeline.

\section{Results}\label{sec:results}
In this section, we present the results of the above pipeline applied
to CMB simulations with varying levels of CS contribution, different beam
  resolutions and various experimental noise levels. Sections~\ref{sec:ideal} and ~\ref{sec:realistic}
investigate noise-free and more realistic cases, respectively.

The curvelet decomposition step of our algorithm (Section~\ref{sec:curvelet}) provides us with four maps, $n_{\rm curvelet}=4$:  the map itself, and the last three (i.e., the fifth to the seventh) curvelet components.
In the ECA step (Section~\ref{sec:ECA}) we use $n_{\rm filter}=5$ filters for edge extraction, corresponding to four differentiation schemes and a case with no filtering at all.
Finally, applying different statistical tools (Section~\ref{sec:stat})
on a single gradient map for a given curvelet component gives $n_{\rm statistics}=5$ measures $\Delta_{\diamond}^{\times}(G\mu)$.  Thus, there are
$n_{\rm curvelet}\times n_{\rm filter}\times n_{\rm statistics}=100$ different combinations
of image-processing and statistical settings to be used in our
pipeline.

In order to quantify the capability of the pipeline in detecting CSs,
we estimate the statistical difference of $\Delta^{\times}(G\mu)$ and
$\Delta^{\times}(G\mu=0)$ (corresponds to a map with no CS
network). The significance of this deviation is systematically checked
by computing the Student's t-test for equal sample sizes and unequal
means and variances as

\begin{widetext}
\begin{eqnarray}
t_{\rm PDF}^{\times}(G\mu)&=&\left[ {\overline {\Delta^{\times}_{\rm PDF}}(G\mu)}-{\overline {\Delta^{\times}_{\rm PDF}}(G\mu=0)}\right]
 \sqrt{\dfrac{N_{\rm sim}}{\left[\sigma^{\times}_{\rm
         PDF}(G\mu)\right]^2+\left[\sigma^{\times}_{\rm
         PDF}(G\mu=0)\right]^2}}\,, \\
t_{\rm TT}^{\times}(G\mu)&=&\left[ {\overline {\Delta^{\times}_{\rm TT}}(G\mu)}-{\overline {\Delta^{\times}_{\rm TT}}(G\mu=0)}\right]
 \sqrt{\dfrac{N_{\rm sim}}{\left[\sigma^{\times}_{\rm
         TT}(G\mu)\right]^2+\left[\sigma^{\times}_{\rm
         TT}(G\mu=0)\right]^2}}\,, \\
t_{\diamond}^{\times}(G\mu)&=&\left[ {\overline {\Delta_{\diamond}^{\times}}(G\mu)}-{\overline {\Delta_{\diamond}^{\times}}(G\mu=0)}\right]
 \sqrt{\dfrac{N_{\rm
       sim}}{\left[\sigma_{\diamond}^{\times}(G\mu)\right]^2 + \left[\sigma_{\diamond}^{\times}(G\mu=0)\right]^2}}\,,
\end{eqnarray}
\end{widetext}
where $\sigma$ is the standard deviation,
$N_{\rm sim}$ is the number of simulations and
$\overline {\Delta^{\times}}$ denotes the mean value of $\Delta^{\times}$ over
the $N_\mathrm{sim}$ maps.

We finally calculate the P-value statistics for
the above Student's t-test and determine the P-value as a function of $G\mu$ for all parameter settings in the pipeline.
In the following, the minimum detectable $G\mu$, denoted by $G\mu_{\rm min}$, refers to the $G\mu$ whose P-value is smaller than a given threshold. We take this threshold to be $0.0027$,
corresponding to the $3\sigma$ frequentist level.
 \subsection{Ideal Case}\label{sec:ideal}
We apply our proposed CS-detection pipeline in all its different
parameter settings, explained in Section~\ref{sec:pipeline}, to
noise-free simulated CMB skies with various levels of CS contributions. Our ideal experiment corresponds to a noise free CMB sky convolved with the beam of
an ACT-like telescope.

More specifically, we construct different gradient maps of various
curvelet components of the GB and GSB maps, and then compare some of
their statistical properties using the measures outlined earlier.
In the following we report, for any given statistical measure,  the lowest $G\mu_{\rm min}$ and its corresponding parameter setting (i.e., curvelet component and ECA filter).  Table~\ref{table:all} summarizes the results.
%
 %
\begin{itemize}
  \item[1)] With $\Delta_{\rm {PDF}}$ as the measure, we find that applying the {\it Laplacian} filter in the ECA step on the seventh curvelet component gives the best detection of the CS signature.
We conclude that, in the absence of instrumental noise and foreground
contamination, our proposed method is basically capable of
robustly detecting the imprints of CSs with string tension
 $G\mu \gtrsim 4.3\times 10^{-10}$ using the $c7$-{\it Laplacian}-$\Delta_{\rm {PDF}}$ sequence. 

\item[2)] With $\Delta_{\rm{TT}}$ as the measure, we find that
  applying the {\it Sobel} filter on the seventh curvelet
  component  results in the best CS detection. This sequence of steps
  is able to discriminate maps with contributions from CSs with
   $G\mu\gtrsim2.3\times 10^{-9}$ from null sky maps. 

\item[3)] With $\Delta_{\rm{pk-pk}}$ as the measure, we find that
  applying the {\it Laplacian} filter on the fifth component  yields the best detection of  the
  CS contribution, setting  the lower detection bound of  $G\mu\gtrsim8.7\times 10^{-10}$. 

\item[4)] With $\Delta_{\rm{up-up}}$ as the measure, we find that the
  seventh curvelet component and the {\it Derivative} filter are most
  sensitive to the CS signature, giving $G\mu\gtrsim8.5\times 10^{-10}$.

\item[5)] Finally with  $\Delta_{\rm{up-pk}}$ as the measure, we
  conclude that applying the {\it Derivative} filter on the seventh
  curvelet component gives the best detection for CSs, able to detect
  strings with tensions $G\mu\gtrsim8.7\times 10^{-10}$. 
\end{itemize}

 In the next section we investigate the performance of the pipeline in
 more realistic scenarios by including both the instrumental noise
   and beam resolution of current and future CMB experiments.


\subsection{Realistic Case}\label{sec:realistic}

 In the previous section, we investigated the performance of our
 proposed CS-detection pipeline for simulations of ideal CMB
 observations.
 Now we take into account noise contamination as explained in Section~\ref{sec:Gnoise}.  It turns out that, unlike the ideal case, the role of the curvelet decomposition becomes less significant as the noise level increases. Also, the scale of the best curvelet component for CS detection depends on the noise level.
 Table~\ref{table:all} summarizes the results of our search for
 $G\mu_{\rm min}$ in these experimental setups, and presents the
 optimum pipeline settings for best CS signal recovery using various
 statistical measures.
 One may note that some experimental setups are
 not present in the table, e.g., the ACT-like case with
 $\Delta_{\rm {TT}}$. This is because for these cases, and for
   a given beam, the minimum detectable value of $G\mu$ is weaker than
   the explored range. For an ACT-like telescope, this corresponds to
   $G\mu_{\rm min} \gtrsim 5 \times 10^{-7}$.

 Our results are
   commented below according to the considered statistical measure.

\begin{itemize}
\item[1)] Using the measure $\Delta_{\rm {PDF}}$.

  For an ACT-like instrumental noise level, CSs are detectable with
tensions $G\mu\gtrsim 1.3 \times 10^{-7}$ with {\it
  Scharr}-filtered maps.
For CMB-S4 phase I- and II-like noise levels, the minimum detectability slightly improves to $G\mu\gtrsim  1.2 \times 10^{-7}$ for the {\it Scharr}-filtered map and the {\it Scharr}-filtered fifth component, respectively.
 For a {\it Planck}-like case one gets $G\mu\gtrsim 4.8 \times 10^{-7}$ with the {\it Scharr}-filtered sixth component.

\item[2)] Using $\Delta_{\rm {TT}}$ as the criterion.

For a CMB-S4 phase I-like noise level, we get the lower bound $G\mu \gtrsim  5.0 \times 10^{-7}$, from the
{\it Sobel}-filtered fifth components.
For a CMB-S4 phase II-like noise level,  the  fifth component itself (with no filtering) yields the
best detectability with $G\mu \gtrsim 4.9 \times 10^{-7}$.
The lower bound for a {\it Planck}-like experiment would be $G\mu \gtrsim 9.4 \times 10^{-7}$, using the $Sobel$-filtered fifth component.

\item[3)] For $\Delta_{\rm {pk-pk}}$, the results are as follows.

For CMB-S4 phase I- and II-like noise levels, using the {\it Sobel}-filtered fifth and sixth components shows that CSs with tensions $G\mu\gtrsim 5.0 \times 10^{-7}$ and $G\mu\gtrsim 4.8 \times 10^{-7}$ are respectively detectable.
While for a {\it Planck}-like experimental setup would reduce this string detectability to $G\mu \gtrsim  8.9 \times 10^{-7}$ obtained through applying the {\it Laplacian} filter on the map itself.

\item[4)] Using $\Delta_{\rm {up-up}}$ as the statistical criterion.

For an ACT-like noise level, the minimum detectability is $G\mu\gtrsim 4.9 \times 10^{-7}$, corresponding to {\it Scharr}-filtered fifth component.
With CMB-S4 phase I- and phase II-like noise levels, using the {\it Sobel}-filters on the map and on the
fifth component yields the minimum
detectability of $G\mu\gtrsim 4.9 \times 10^{-7}$ and $G\mu\gtrsim 2.4 \times 10^{-7}$, respectively.
The minimum detectability corresponding to the {\it Planck}-like observational scenario in this case is provided by the $Sobel$-filtered fifth component to be $G\mu\gtrsim 8.4 \times 10^{-7}$.

\item[5)] Using the measure $\Delta_{\rm {up-pk}}$.

For the CMB-S4 phase I-like noise level, using the {\it Scharr}-filtered map gives $G\mu\gtrsim 5.0 \times 10^{-7}$, improving to
$G\mu\gtrsim  2.4 \times 10^{-7}$ for the CMB-S4 phase II-like noise level with the use of the {\it Sobel}-filtered fifth component.
\end{itemize}
 \begin{table}
\centering
\caption{Lowest detectable $G\mu$, labeled as $G\mu_{\rm min}$, of the
  CSs network superimposed on the CMB map, using various statistical
  measures. The first column, "Measure", contains the probability density
  function (PDF), the correlation function of temperature fluctuations
  versus angle separation (TT), the unweighted TPCF of local maxima
  (pk-pk), the unweighted TPCF of up-crossings (up-up) and the unweighted
  cross-correlation of up-crossings and peaks (up-pk).  The "Method"  corresponds
  to the sequence of curvelet and ECA filters leading to the best CS detection for the given statistical measure.
The "Map" and "FWHM"  characterize the experimental setup.  CMB-S4-like (I) and (II) represent the phases I and II of a CMB-S4-like experiment, respectively. }
    \label{table:all}
     \begin{tabular}{l|cccccl}
\hline
Measure&Method &   Map & FWHM&$G\mu_{\rm min}$  \\
\hline\hline
&c7-lap  &   No noise &      $0.9^{'}$ &  $4.3 \times 10^{-10}$ \\
&c5-sch  &  CMB-S4-like (II) &      $0.9^{'}$ &   $1.2 \times 10^{-7}$ \\
PDF&map-sch &   CMB-S4-like (I) &      $0.9^{'}$ &   $1.2 \times 10^{-7}$ \\
& map-sch &        ACT-like &   $0.9^{'}$ &   $1.3 \times 10^{-7}$ \\
&c6-sch &        {\it Planck} &      $5^{'}$ &   $4.8 \times 10^{-7}$ \\
\hline\hline
&c7-sob  &   No noise &      $0.9^{'}$&  $2.3 \times 10^{-9}$ \\
&c5-none &  CMB-S4-like (II) &      $0.9^{'}$&  $4.9 \times 10^{-7}$ \\
TT&c5-sob  &   CMB-S4-like (I) &      $0.9^{'}$&  $5.0 \times 10^{-7}$ \\

&c5-sob &        {\it Planck} &      $5^{'}$ &   $9.4 \times 10^{-7}$ \\
\hline\hline
&c5-lap  &   No noise &      $0.9^{'}$&  $8.7 \times 10^{-10}$ \\
&c6-sob  &  CMB-S4-like (II) &      $0.9^{'}$&   $4.8 \times 10^{-7}$ \\
pk-pk&c5-sob  &   CMB-S4-like (I) &      $0.9^{'}$&   $5.0 \times 10^{-7}$ \\
&map-lap &        {\it Planck} &      $5^{'}$ &   $8.9 \times 10^{-7}$ \\
\hline\hline
&c7-der  &   No noise &      $0.9^{'}$&  $8.5 \times 10^{-10}$ \\
&c5-sob  &  CMB-S4-like (II) &      $0.9^{'}$&   $2.4 \times 10^{-7}$ \\
up-up&map-sob &   CMB-S4-like (I) &      $0.9^{'}$&   $4.9 \times 10^{-7}$ \\
&c5-sch  &        ACT-like &     $0.9^{'}$ &   $4.9 \times 10^{-7}$ \\
&c5-sob &        {\it Planck} &      $5^{'}$ &   $8.4 \times 10^{-7}$ \\
\hline\hline
&c7-der  &   No noise &      $0.9^{'}$&  $8.7 \times 10^{-10}$ \\
up-pk&c5-sob  &  CMB-S4-like (II) &      $0.9^{'}$&   $2.4 \times 10^{-7}$ \\
&map-sch &   CMB-S4 (I)-like &      $0.9^{'}$&   $5.0 \times 10^{-7}$ \\
\hline
\end{tabular}
\end{table}


\section{Summary and Conclusions}
Increasing the quantity and quality of observational data provides
the opportunity to search for possible features present in
beyond-the-standard models. A well-motivated example is the CS
network, possibly produced in a series of symmetry breaking phase
transitions in the very early Universe. If such a network exists,
CMB anisotropies are among the powerful observational data sets for
their discovery.  However, finite instrumental noise and beam
smearing effect greatly reduce the detectability of their trace.

Our purpose in this work was to exploit the specific anisotropy patterns, especially  the line-like discontinuities,  induced by the CSs on the CMB temperature maps, to enhance the string network detectability.
Therefore, we have tested a multi-step pipeline which employs
image-processing tools to amplify the string signal as well as
statistical measures to quantify the deviation of the simulated data from pure Gaussian inflation-induced anisotropies.

The first image-processing step is a curvelet decomposition, an
appropriate tool for the detection of elongated sharp edges. As a
result, it isolates the components with the highest string
contribution. The second image-processing step is based on the
extended Canny algorithm, and produces gradient (or filtered) maps
with magnified string signatures. The gradient maps are then passed to
the statistical unit of the pipeline to investigate the
detectability of the strings with different $G\mu$ values, thereby
enabling us to compare the efficiency of the various settings in the pipeline.
These settings, or degrees of freedom,
correspond to the
various available options for curvelet decomposition, ECA filtering and
statistical measures. The
pipeline  explores the space of these parameters and finds the  setting which best constrains the
contribution of the string network to the CMB anisotropies. This
parameter set depends on the experiment characterization, such as beam
and noise level.

We have found that, the curvelet components describing the smaller
scales are preferred by the algorithm. This is expected given the
small scale nature of the kicks produced by the CSs in the data. In our
analysis, this small scale mode corresponds to the seventh mode for an
ideal experiment.
In the presence of instrumental noise, however, the scale where the
CS signal dominates the small-scale noise contamination depends on
the noise level. Therefore, the best curvelet component also varies according to
the experimental set-up. It also turns out that the
two-point statistics, being relatively powerful for an ideal
experiment, get highly contaminated by instrumental noise. In
  these situations, the PDF is the preferred statistical measure
while the results end up being relatively insensitive to the choice
of the filter.

For the most efficient pipelines, we could detect the imprints of CS network with
tensions as low as $G\mu_{\rm min} = 4.3 \times 10^{-10}$ for a
noise-free experiment having a beam of $\mathrm{FWHM=0.9'}$. The
sequence of pipeline parameter for this case is the seventh curvelet
component, the {\it Laplacian} filter and the one point PDF as the
statistical measure (or, $c7-Laplacian$-PDF).
Including more realistic noise levels increases this minimum detectable tension to $G\mu_{\rm min}=1.3\times 10^{-7}$ and $G\mu_{\rm min}= 1.2\times 10^{-7}$ for ACT-like and CMB-S4-like noise levels, respectively, both with a beam pattern described in Section~\ref{sec:Gnoise}.
All results are listed in Table \ref{table:all}.

We have also considered  the unweighted TPCF of up-pk. For noise-free
  case, we have found $G\mu\gtrsim 8.7\times 10^{-10}$ for $c7-Derivative$
method.
  Based on the CMB-S4 phase II observational strategy, this type of cross-correlation
resulted in $G\mu\gtrsim 2.4\times 10^{-7}$ for
$c5-Sobel$. 

As a final remark, let us notice that the free parameters
selected in this work are by no means exclusive, and should be merely
considered as starting points.
An obvious extension of the work would
thus be to include other statistical measures, which could be possibly more
sensitive to CS imprints, and explore other filters to assess their
performance in edge detection.
One could use deep-learning approaches toward systematic decision making to choose most sensitive features for the CS network detection. This work is in progress. Another possible improvement on our
work would be to perform Bayesian model comparison between the best
pipelines obtained here on
$G\mu$ on real data in order to extract the tightest bound.

\section*{Acknowledgements}

The numerical simulations were carried out on Baobab at the computing
cluster of the University of Geneva.

\bibliography{bib}{}
\bibliographystyle{mnras}

\bsp    
\label{lastpage}
\end{document}